\documentclass[superscriptaddress,twocolumn,10pt]{revtex4-1}

\usepackage{amsmath}    
\usepackage{graphicx}   
\usepackage{verbatim}   
\usepackage{color}      
\usepackage{subfigure}  
\usepackage{hyperref}   
\usepackage{amssymb}
\usepackage{amsthm}
\usepackage{amsfonts}
\usepackage{float} 
\begin{document}


\title{Effect of Pore Geometry on the Compressibility \\of a Confined Simple Fluid}

\author{Christopher D. Dobrzanski}
\author{Max A. Maximov}
\author{Gennady Y. Gor} \email[Corresponding author, e-mail: ]{gor@njit.edu} \homepage[\\ URL: ]
{http://porousmaterials.net}
\affiliation{Otto H. York Department of Chemical, Biological, and Pharmaceutical Engineering, New Jersey Institute of Technology, University Heights, Newark, NJ 07102, USA}

\date{\today}

\begin{abstract} 
Fluids confined in nanopores exhibit properties different from the properties of the same fluids in bulk, among these properties are the isothermal compressibility or elastic modulus. The modulus of a fluid in nanopores can be extracted from ultrasonic experiments or calculated from molecular simulations. Using Monte Carlo simulations in the grand canonical ensemble, we calculated the modulus for liquid argon at its normal boiling point (87.3~K) adsorbed in model silica pores of two different morphologies and various sizes. For spherical pores, for all the pore sizes (diameters) exceeding 2~nm, we obtained a logarithmic dependence of fluid modulus on the vapor pressure. Calculation of the modulus at saturation showed that the modulus of the fluid in spherical pores is a linear function of the reciprocal pore size. The calculation of the modulus of the fluid in cylindrical pores appeared too scattered to make quantitative conclusions. We performed additional simulations at higher temperature (119.6~K), at which Monte Carlo insertions and removals become more efficient. The results of the simulations at higher temperature confirmed both regularities for cylindrical pores and showed quantitative difference between the fluid moduli in pores of different geometries. Both of the observed regularities for the modulus stem from the Tait-Murnaghan equation applied to the confined fluid. Our results, along with the development of the effective medium theories for nanoporous media, set the groundwork for analysis of the experimentally-measured elastic properties of fluid-saturated nanoporous materials. 
\end{abstract}

\maketitle

\section{Introduction}
\label{sec:Intro}

The thermodynamic properties of a confined fluid differ from that of a fluid in bulk at the same temperature and pressure \cite{Gelb1999, Huber2015}. Ultrasonic experiments on fluid-saturated nanoporous materials provide the way to probe one of those thermodynamic properties: compressibility or elastic (hydrostatic) modulus of the confined fluid \cite{Page1995, Schappert2014, Gor2018Gassmann}. Although the first ultrasonic measurements on fluid-saturated nanoporous samples have been carried out in the early 1980s \cite{Murphy1982}, there have been relatively few studies of this kind since then. Warner and Beamish used ultrasonic experiments to investigate the surface area of nanoporous materials \cite{Warner1988}. Page et al. studied pore-space correlations with adsorption on Vycor glass and effects of pore connectivity and were the first to report the elastic modulus of confined fluid (n-hexane) \cite{Page1993, Page1995}. Many works have employed ultrasonic experiments to study phase transitions of confined fluids \cite{Molz1993, Molz1995, Beaudoin1996, Charnaya2001, Schappert2008, Matsumoto2009, Borisov2006, Borisov2009, Charnaya2008}, but did not quantify the elastic properties of confined phases. 

Recently Schappert and Pelster used ultrasonic measurements to study the changes of elastic properties of fluid and solid phases of argon, nitrogen, and oxygen confined in nanoporous materials at low temperatures \cite{Schappert2013JoP, Schappert2013N2, Schappert2014, Schappert2014Langmuir, Schappert2016JPCC}. Since argon is one of the simplest systems for molecular simulations, these works stimulated the development of macroscopic \cite{Gor2014}, and molecular modeling approach to the calculation of elastic properties of confined fluids \cite{Gor2015compr, Gor2016Tait, Gor2017Biot}.

Refs.~\onlinecite{Gor2015compr, Gor2016Tait, Gor2017Biot} presented the calculations of the elastic modulus of argon confined in spherical silica pores. The model used in those calculations is suitable to represent many nanoporous materials, such as SBA-16 silica \cite{Zhao1998}, 3DOm carbon \cite{Fan2008}, KLE and SLN-326 silica \cite{Rasmussen2010}. The experimental data available in the literature is mainly for Vycor glass \cite{Warner1988, Page1995, Schappert2014}, which has different morphology. Pores in Vycor form a network of interconnected channels \cite{Levitz1991}. Since the length of these pores significantly exceeds its diameter and the diameter does not vary much along the length of the pores, the behavior of fluids in Vycor glass is often simulated in a cylindrical pore model \cite{Landers2013}. Thus we expect that simulations of argon in cylindrical pores would be a more rigorous representation of the experimental system studies in Refs.~\onlinecite{Warner1988, Page1995, Schappert2014}.

Note that the Refs.~\onlinecite{Gor2015compr, Gor2016Tait, Gor2017Biot} are not the only theoretical works studying the compressibility of confined fluids. Rickman used conventional Metropolis Monte Carlo simulations and stress correlation functions to determine the elastic properties of Lennard-Jones (LJ) fluid in a slit pore \cite{Rickman2012}. Sun and Kang \cite{Sun2014} and Keshavarzi et al. \cite{Keshavarzi2016} employed density functional theory to determine the elastic properties of LJ fluid in spherical and slit pores respectively. Vadakkepatt and Martini investigated the compressibility of fluids confined in slit pores using molecular dynamics simulations \cite{Vadakkepatt2011, Martini2010}. However, none of these works calculated the moduli in the context of adsorption experiments and ultrasonics.  

The primary goal of the current paper is to investigate the elastic properties of a simple fluid in cylindrical confinement, which is assumed to be a more realistic representation of the system used in ultrasonic experiments by several groups \cite{Warner1988, Page1995, Schappert2014}. Here we consider the same system and use the same methods as in \cite{Gor2015compr}: we model argon at its normal boiling temperature confined in silica mesopores using conventional grand canonical Monte Carlo (GCMC) simulations \cite{Norman1969}. However, while Ref.~\cite{Gor2015compr} dealt exclusively with the spherical pore model, here we consider both spherical and cylindrical pore models. Therefore, we investigate the effect of the pore shape on the elastic properties of confined fluids and examine the validity of the relations between modulus and pore size (diameter) \cite{Gor2015compr} and between modulus and pressure \cite{Gor2016Tait} for the cylindrical pore model.

\section{Methods}
\label{sec:Methods}

\subsection*{Compressibility and Bulk Modulus}

In thermodynamics, the elastic properties are typically presented in terms of the isothermal compressibility $\beta_T$. We start from introducing equations for compressibility, but the results of the calculations are more convenient to represent in the form of the isothermal elastic modulus $K_T = \beta_T^{-1}$ which is more relevant to ultrasonics.

For a macroscopic system, the isothermal compressibility $\beta_T$ is defined as
\begin{equation}
\label{beta-def}
\beta_T \equiv - \frac{1}{V} \left( \frac{\partial V}{\partial P} \right)_{N,T}
\end{equation}
where $V$ is the system volume, $P$ is the fluid pressure, and $T$ is the absolute temperature. Here, following Refs.~\onlinecite{Landau5, Gor2015compr, Gor2016Tait}, we use the same definition of $\beta_T$ for the fluid confined in the pore. We determine the overall fluid compressibility in the pore which corresponds to the macroscopic average compressibility that can be extracted from the experimental data on fluid-saturated porous samples using effective medium analysis. 

\subsection*{Compressibility by Statistical Mechanics}

Classical statistical mechanics allows for the calculation of the compressibility of the fluid in the pore from the fluctuations in the number of particles in the pore $N$ in the grand canonical ensemble through the following relation \cite{Coasne2009}
\begin{equation}
\label{beta-fluct}
\beta_T = \frac{V\langle \delta N^2\rangle}{k_{\rm{B}}T\langle N\rangle ^2} 
\end{equation}
where $\langle \delta N^2\rangle$ is the variance of $N$ and $k_{\rm{B}}$ is the Boltzmann constant. Eq. \ref{beta-fluct} can be applied to a small system as long as the fluctuations obey a Gaussian distribution \cite{Landau5, Gor2015compr}. Thus, molecular simulation of a fluid in the pore performed in the grand canonical ensemble can provide data for calculation of $\beta_T$.

\subsection*{Compressibility by Macroscopic Thermodynamics}

Another derivation of the compressibility of a confined fluid was done by one of us \cite{Gor2014} from the same starting point, Eq. \ref{beta-def}. By neglecting the anisotropy of pressure and considering only a macroscopic average, the pressure $P$ in the pore, which is also known as the solvation pressure, can be determined from the grand thermodynamic potential $\Omega$ \cite{Ravikovitch2006, Gor2010}
\begin{equation}
\label{pressure}
P = -\left( \frac{\partial \Omega}{\partial V} \right)_{\mu , T}.
\end{equation}
Also, the pressure in the pore $P$ is related to the chemical potential $\mu$ of the fluid via the Gibbs-Duhem equation
\begin{equation}
\label{gibbsduhem}
dP = nd\mu
\end{equation}
where $n$ is the average particle density in the pore defined as $n \equiv N/V$.

Assuming that the number of particles in the pore and the temperature are constant, Eq. \ref{gibbsduhem} can be used to rewrite Eq. \ref{beta-def} as
\begin{equation}
\label{beta-thermo}
\beta_T = \frac{1}{n^2}\left( \frac{\partial n}{\partial \mu} \right)_{N, T}.
\end{equation}
Since, at constant temperature and when Eq. \ref{gibbsduhem} is valid, Eq. \ref{beta-thermo} is only a function of intensive variables (i.e. it does not depend on $N$ nor $V$), we can write
\begin{equation}
\label{partial-equality}
\left( \frac{\partial n}{\partial \mu} \right)_{N, T} = \left( \frac{\partial n}{\partial \mu} \right)_{V, T}.
\end{equation}
This transformation is important because in the grand canonical ensemble, the number of particles does indeed change while the volume of the system is kept constant.

Since the vapor pressure is low, the vapor can be considered an ideal gas. Then, the chemical potential is related to the vapor pressure in equilibrium with the fluid in the pore by the relation
\begin{equation}
\label{chem-pot}
\mu = k_{\rm{B}}T\ln(p/p_0) + \mu_0(T)
\end{equation}
where $p_0$ and $\mu_0(T)$ are the vapor pressure and chemical potential at saturation respectively. Eq. \ref{beta-thermo} can be rewritten using Eqs. \ref{partial-equality} and \ref{chem-pot} as \cite{Gor2014}
\begin{equation}
\label{beta-thermo-final}
\beta_T = \frac{1}{n^2} \frac{p/p_0}{k_{\rm{B}}T}\left( \frac{\partial n}{\partial (p/p_0)} \right)_{V, T}.
\end{equation}
Therefore, to calculate the compressibility of a confined fluid using the thermodynamic method, one only needs the density $n$ of the fluid in the pore as a function of the relative pressure $p/p_0$ which is known as the adsorption isotherm. The derivative in Eq. \ref{beta-thermo-final} can be obtained from the slope of the isotherm.

\subsection*{Grand Canonical Monte Carlo Simulations}

\begin{figure}[H]
\centering
\includegraphics[width=0.7\linewidth]{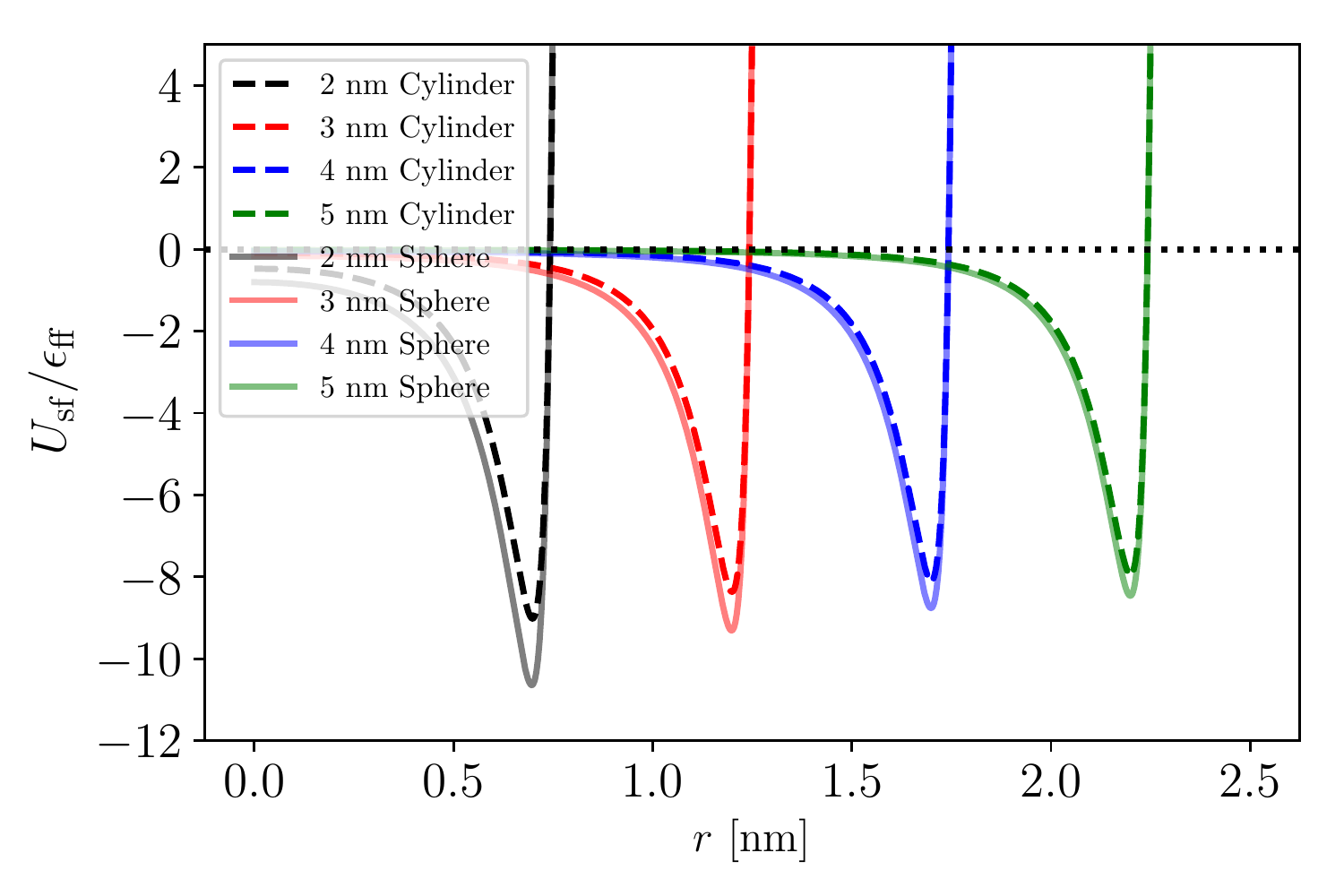}
\caption{The solid-fluid potential for spherical and cylindrical pores. The potential from each individual wall atom is integrated over the surface of the pore. For each of the pore sizes, the deeper potential of the spherical pore is consistent with the higher degree of confinement due to the closer interactions with the pore walls. The point at which the potential is zero corresponds to the distance from the center of the pore to the center of the outermost fluid atoms, from which one can determine the internal diameter given by Eq.~\ref{dint}.}
\label{fig:Usf} 
\end{figure}

The fluid used for our simulations was argon; interactions between argon atoms were modeled by LJ pair potentials.  We considered two temperatures: the normal boiling point $T = 87.3$~K, which is typical for argon adsorption experiments and close to the temperature in the ultrasonic experiments in Ref.~\onlinecite{Schappert2014}, and at $T = 119.6$~K, which corresponds to the reduced temperature $T^* = 1$. The reason for using this higher temperature in the simulations is discussed below. The simulations were performed at LJ reduced chemical potentials $\mu^{*} = \mu /\epsilon_{\rm{ff}}$ ranging from $-15.0$ to $-9.6$ for 87.3~K and $-23.0$ to $-11.6$ for 119.6~K; the upper limits of $\mu^*$ correspond to the saturation pressure $p_0$ of the fluid. The parameter $\epsilon_{\rm{ff}}$ along with other Lennard-Jones parameters and physical properties used in the simulations are summarized in Table \ref{tab:tabLJ}. The simulations were performed using the conventional GCMC method \cite{Norman1969} based on the Metropolis algorithm \cite{Metropolis1953}. The adsorptive potentials between the fluid atoms and the pore wall were modeled by spherically \cite{Baksh1991} or cylindrically \cite{Tjatjopoulos1988} integrated, site-averaged interaction potentials. For the cylindrical pore model, we used a pore length of $40 \sigma_{\rm{ff}}$ and applied periodic boundary conditions along the direction of the cylinder axis. For the 2~nm pore, we used a length of $80 \sigma_{\rm{ff}}$ to even further increase the number of atoms in the system. Figure \ref{fig:Usf} shows the calculated solid-fluid interaction potentials for spherical and cylindrical pores of the 2, 3, 4, and 5~nm pore sizes that were used in simulations. For each pore size, the potential for the sphere has a deeper well which corresponds to the higher degree of attraction between the fluid and the wall in the sphere than in the cylinder. 

\begin{table}[ht]
\begin{tabular}{|c|c|c|c|c|c|}
  \hline
  Interaction & $\sigma$, nm & $\epsilon/k_{\rm{B}}$, K & $\rho_{\rm{s}}$, nm$^{-2}$ & $r_{\rm{cut}}, \sigma_{\rm{ff}}$ & Ref\\
  \hline
  Ar-Ar & 0.34 & 119.6 & - & 5 & \cite{Vishnyakov2001}\\
  Silica-Ar & 0.30 & 171.24 & 15.3 & 10 & \cite{Ravikovitch1997}\\
  \hline
\end{tabular}
\caption{Parameters for the fluid-fluid (ff) and solid-fluid (sf) interactions for the argon-silica system. $\sigma$ is the LJ diameter, $\epsilon$ is the LJ energy, $\rho_{\rm{s}}$ is the number density of solid LJ sites on the surface, and $r_{\rm{cut}}$ is the cut-off distance where interactions were truncated; no tail corrections were used.} 
\label{tab:tabLJ}
\end{table}

Simulations were done with pore sizes ranging from 2~nm to 6~nm. The pore size refers to the external diameter $d_{\rm{ext}}$ which is taken as the center-to-center distance from one pore wall molecule to the molecule on the opposite side of the pore (see Figure~\ref{fig:dint}). The volume of the pore that is accessible to the fluid atoms $V$ is different from the volume calculated using the external diameter of the pore. To calculate the internal diameter $d_{\rm{int}}$ we used the approach from Refs.~\cite{Rasmussen2010, Gor2012} and extended it to cylindrical geometry (see Appendix), which gives
\begin{equation}
\label{dint}
d_\mathrm{int}\approx d_\mathrm{ext} - 1.7168\sigma_\mathrm{sf}+\sigma_\mathrm{ff}.
\end{equation}

At each pore size and each chemical potential, simulations were run for at least $5 \times 10^9$ trial Monte Carlo moves. Each simulated data point was first equilibrated with at least $10^9$ trial moves that were not considered in calculations. The reduced chemical potential $\mu^*$ was mapped to the relative vapor pressure $p/p_0$ using the Johnson et al. equation of state \cite{Johnson1993}, from which we calculated the reduced chemical potential at vapor-liquid equilibrium to be $\mu^* = -9.6$ at $T=87.3$~K and $\mu^* = -11.6$ at $T = 119.6$~K. Considering the vapor to be an ideal gas, we calculated the pressures at other values of chemical potential (Eq. \ref{chem-pot}).

\section{Results}
\label{sec:Results}

\begin{figure}[H]
\centering
\includegraphics[width=0.7\linewidth]{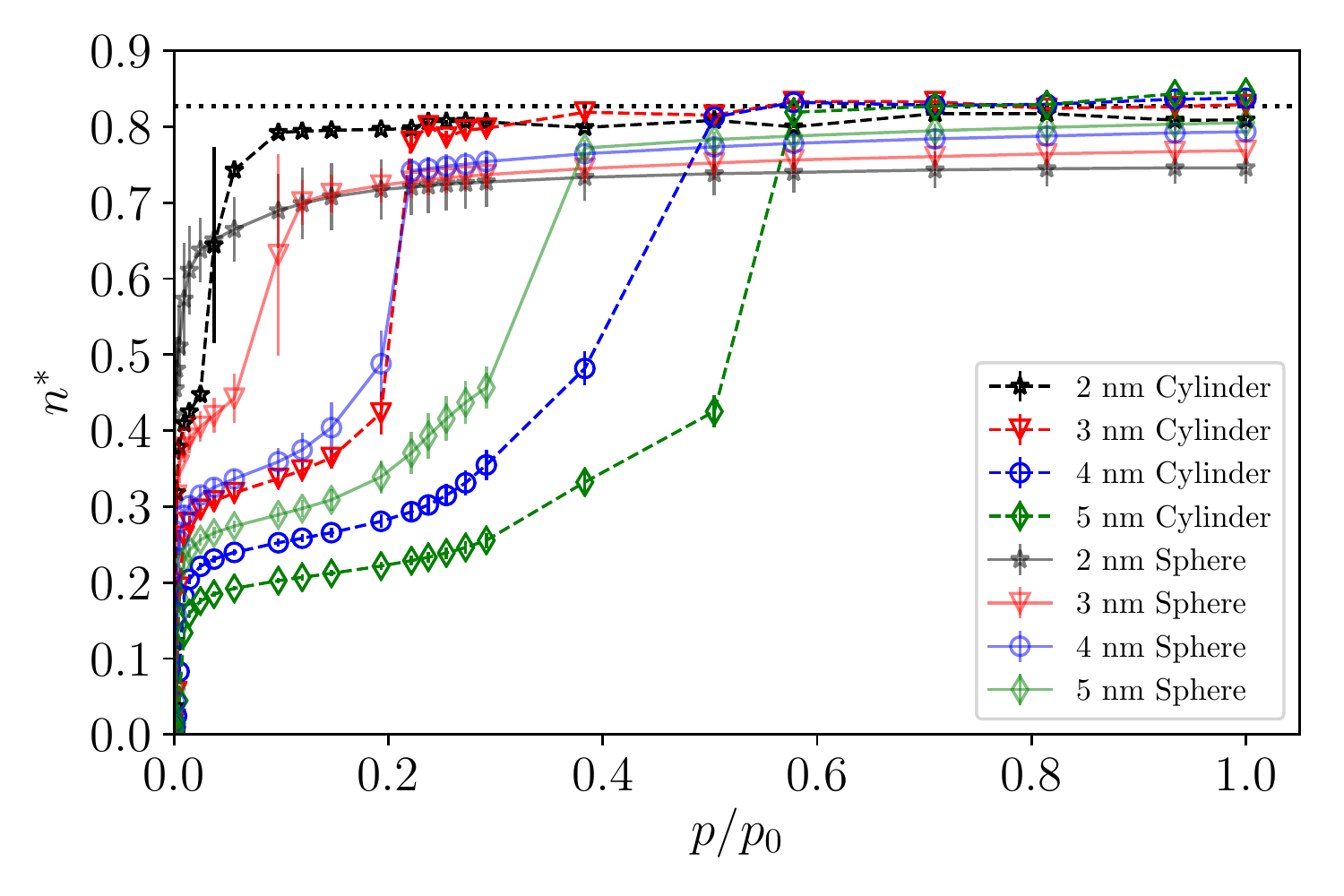}
\includegraphics[width=0.7\linewidth]{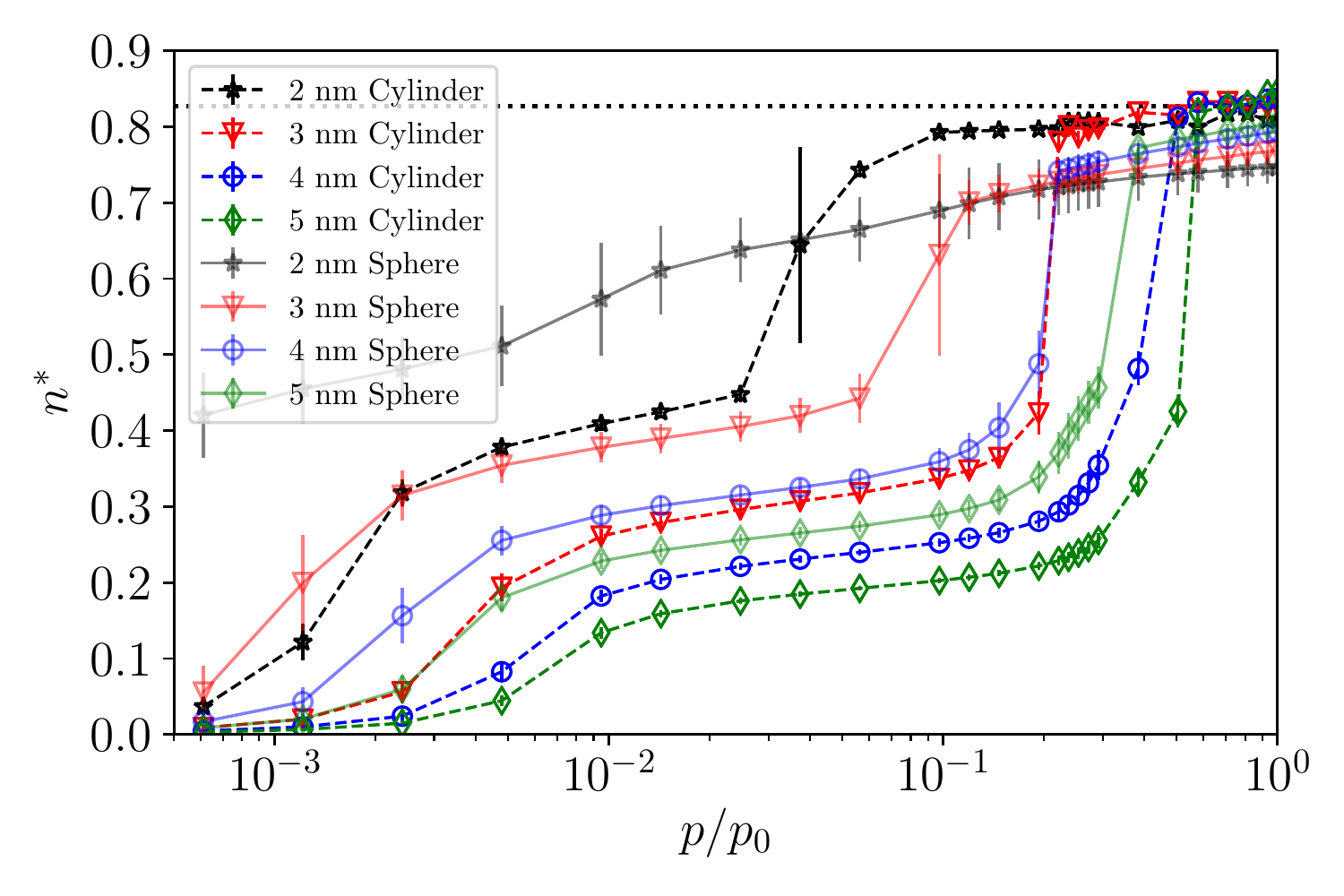}
\caption{GCMC adsorption isotherms for argon at 87.3~K in spherical and cylindrical pores shown as the average reduced fluid density $n^* = N\sigma_\mathrm{ff}^3/V$ plotted versus relative pressure. The top plot is in linear scale and the bottom has pressure in log scale. The horizontal dotted line at $n^* = 0.827$ represents the bulk density. Error bars represent a twice standard deviation error in the fluid density in the pore.}
\label{fig:Isotherms} 
\end{figure}

We constructed GCMC adsorption isotherms from simulations of various pore sizes for both spherical and cylindrical pores at $T=87.3$~K. The complete adsorption isotherms for spherical and cylindrical pores of 2, 3, 4 and 5~nm in size are shown in Figure \ref{fig:Isotherms}. These isotherms display the typical behavior of monolayer formation at very low relative pressures, followed by multilayer formation, after which the pores are rapidly filled via capillary condensation. The spherical pores exhibit capillary condensation at lower pressures than the cylindrical pores, e.g. for 3~nm pores the capillary condensation takes place at $p/p_0 \simeq 0.1$, while for cylindrical pore at $p/p_0 \simeq 0.2$. This suggests that the confinement effects in spherical pores are stronger than in cylindrical, which is consistent with the deeper potential wells in spheres for the $U_{\rm{sf}}$ potential shown in Figure~\ref{fig:Usf}. The simulation data for the cylindrical pore of 3~nm size show another interesting feature: soon after the capillary condensation there is a second step on the adsorption isotherm, corresponding to small but noticeable densification. A similar feature has been recently reported by Siderius et al. for the simulation of Lennard-Jones methane in cylindrical pores using transition-matrix Monte Carlo simulations and attributed to a phase-transition to a more ordered phase \cite{Siderius2017}.

The error bars in Figure \ref{fig:Isotherms} are twice the standard deviation error, related to the fluctuation or variance of the number of atoms $N$ in the pores. The variance in the number of atoms is proportional to the compressibility of the fluid in the pore by Eq. \ref{beta-fluct}; therefore, the data shown in Figure~\ref{fig:Isotherms} can be used for calculation of the compressibility (or elastic modulus of the fluid).

\begin{figure}[H]
\centering
\includegraphics[width=0.7\linewidth]{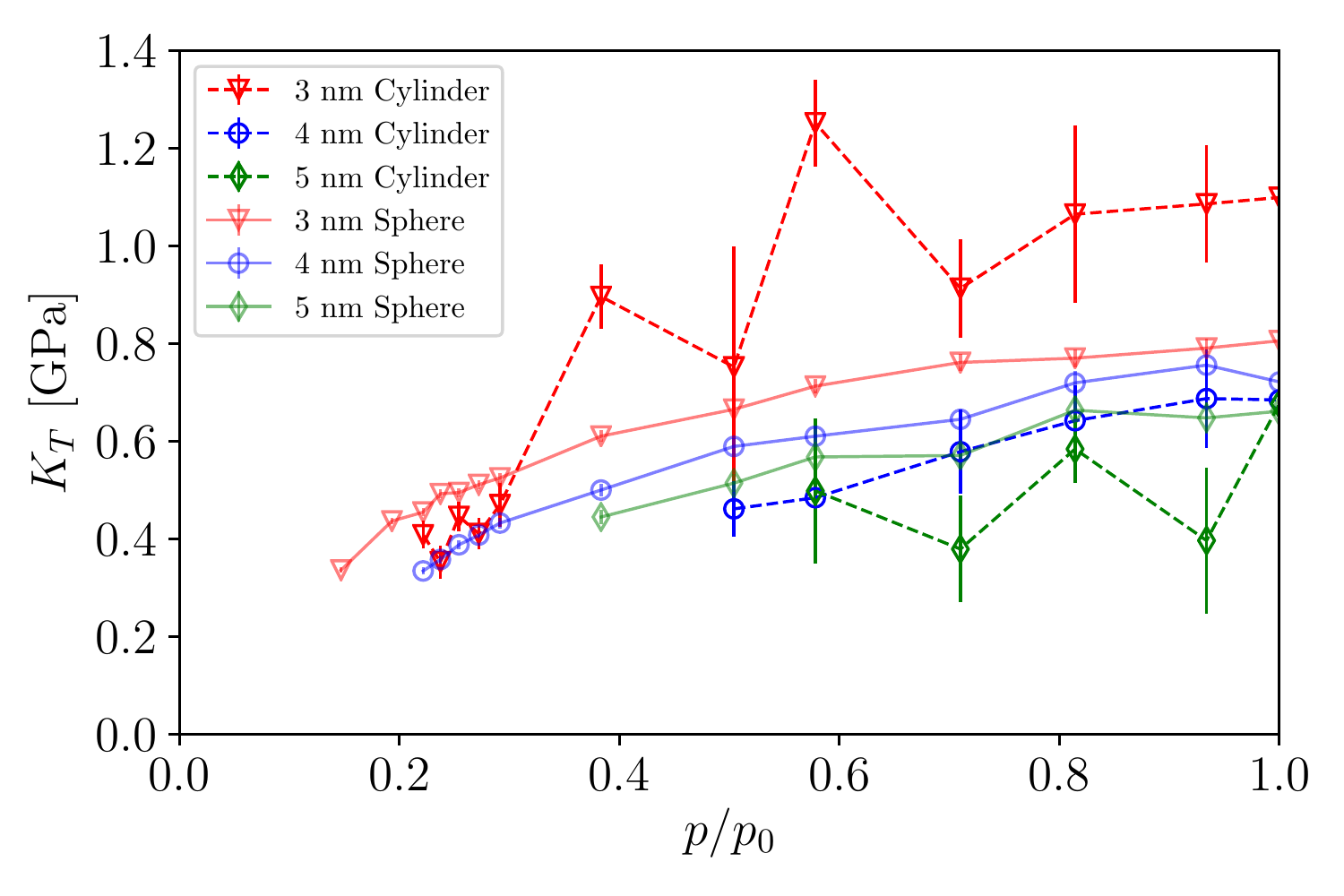}
\includegraphics[width=0.7\linewidth]{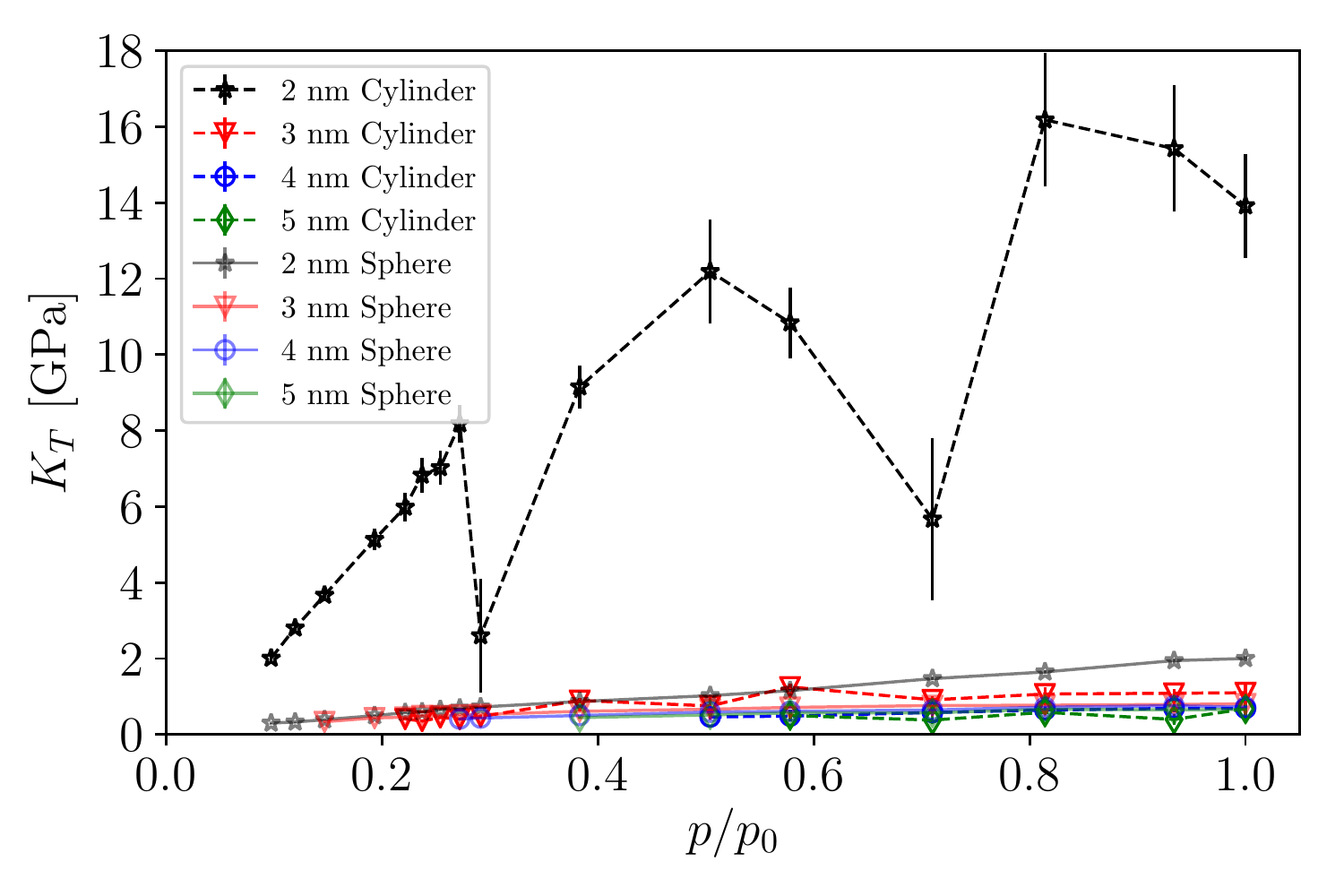}
\caption{(Top) Isothermal modulus $K_T$ of argon in spherical and cylindrical pores of 3, 4, and 5~nm at $87.3$~K calculated using Eq. \ref{beta-fluct} from GCMC simulations as a function of reduced pressure. (Bottom) The same data shown together with the calculations for the 2~nm pore. The error bars represent the correlation error estimated by the method described in Ref.~\onlinecite{Gor2015compr}.}
\label{fig:K-all-points} 
\end{figure}

We calculated the isothermal elastic modulus of the fluid in the pores based on Eq. \ref{beta-fluct}, which is relevant for the pressures above the capillary condensation, when the pores are filled with a liquid-like condensate. The modulus $K_T$ as a function of reduced vapor pressure $p/p_0$ is shown in Figure \ref{fig:K-all-points} for filled pores of 2, 3, 4 and 5~nm size and of both spherical and cylindrical geometry. The upper panel shows the results for the 3, 4, and 5~nm pores. The data for spherical pores show a clear trend: the modulus of fluid exhibits monotonic increase as a function of reduced vapor pressure $p/p_0$. The data for the cylindrical pores are so scattered that it is hard to make a similar conclusion; yet the values of moduli are of the same order of magnitude. The lower panel shows the same data along with the results of the modulus of argon in the 2~nm pore, which exceeds the moduli for the fluid in larger pores by an order of magnitude. Note that the methods used here for the calculation of the elastic moduli are applicable only when the pores are filled with capillary condensate. Even if one could come up with a method to calculate the modulus of the adsorbed fluid at pressures below capillary condensation, it could not be accessed via ultrasonic experiments \cite{Gor2018Gassmann}.

\begin{figure}[H]
\centering
\includegraphics[width=0.45\linewidth]{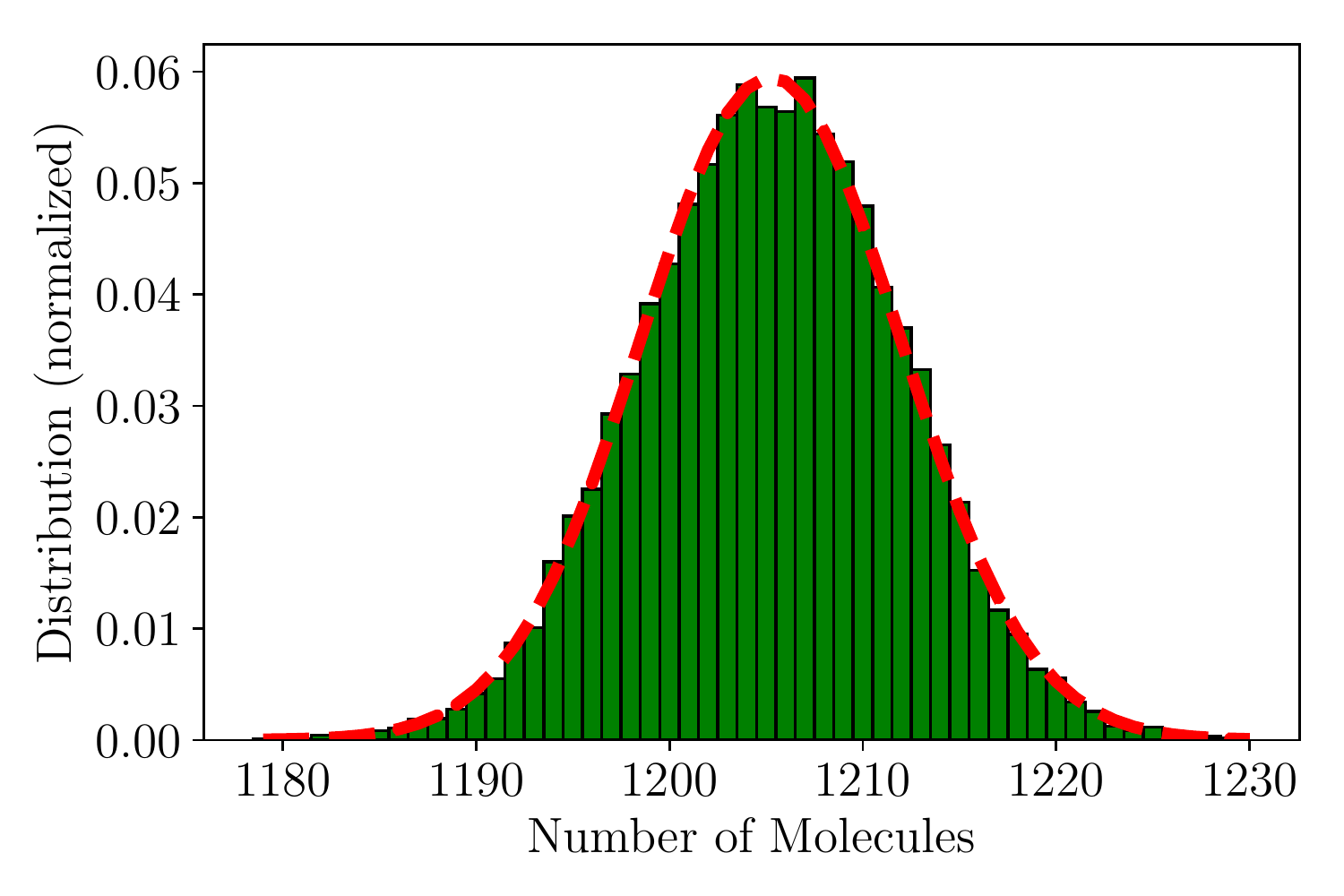}
\includegraphics[width=0.45\linewidth]{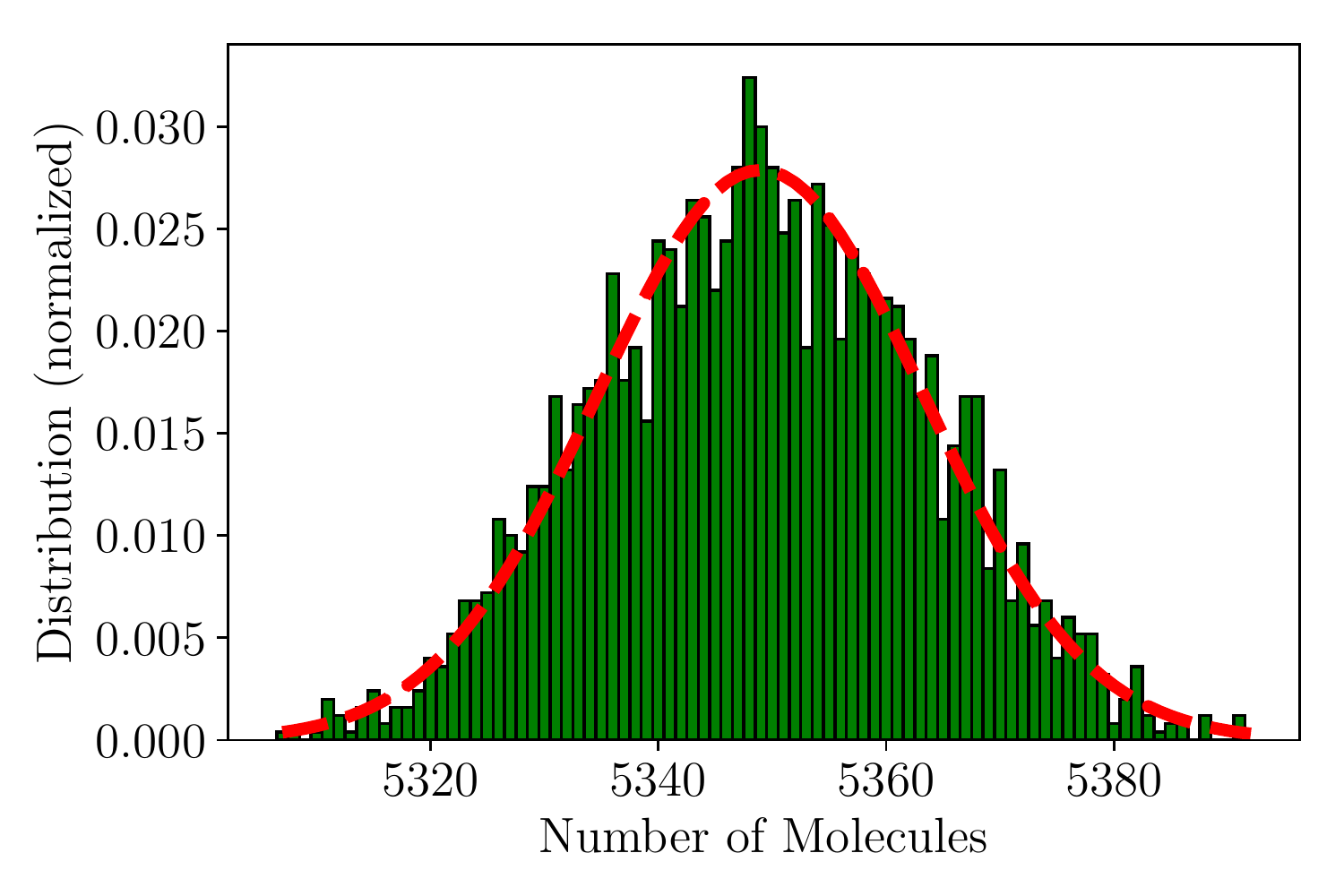} \\
\includegraphics[width=0.45\linewidth]{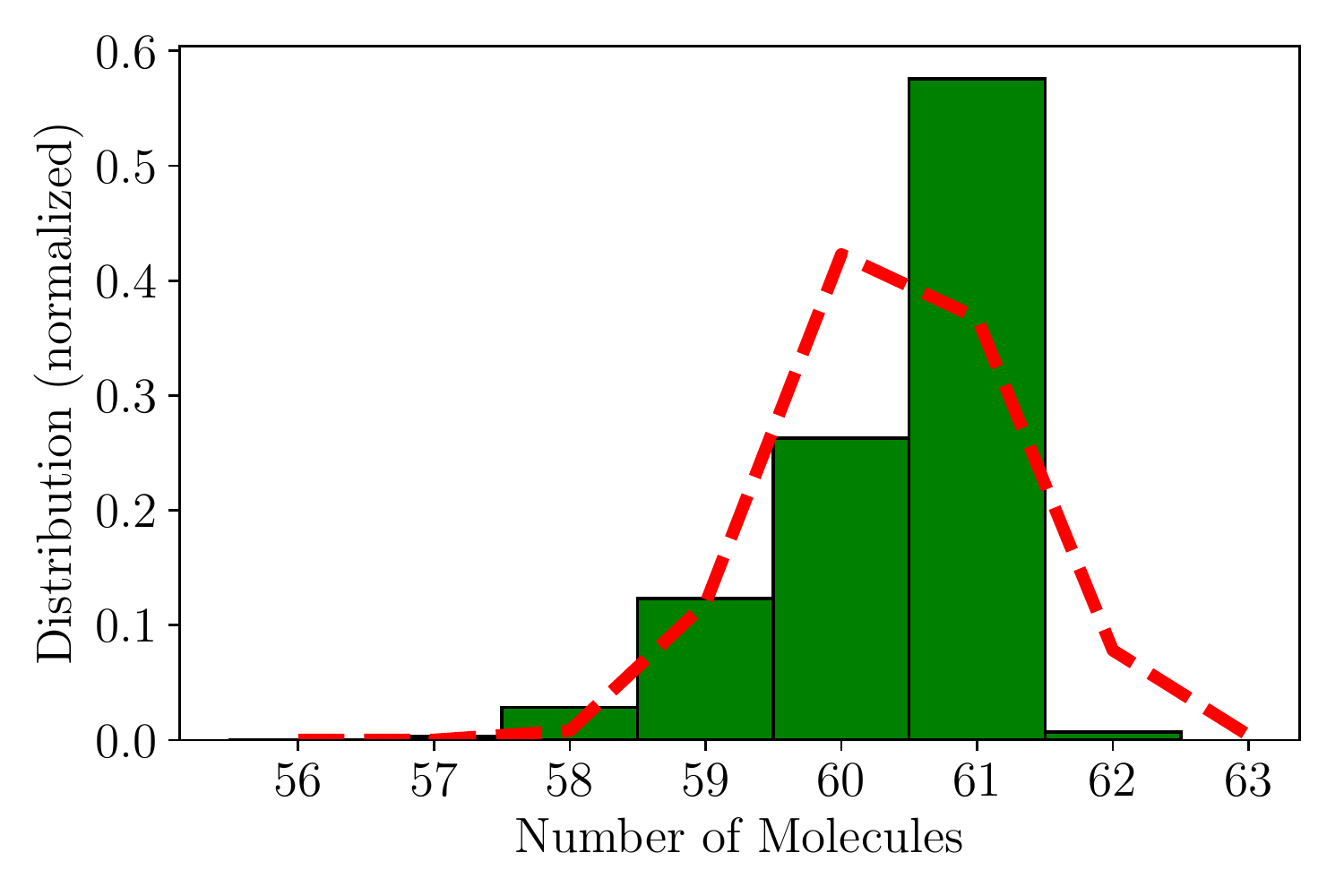}
\includegraphics[width=0.45\linewidth]{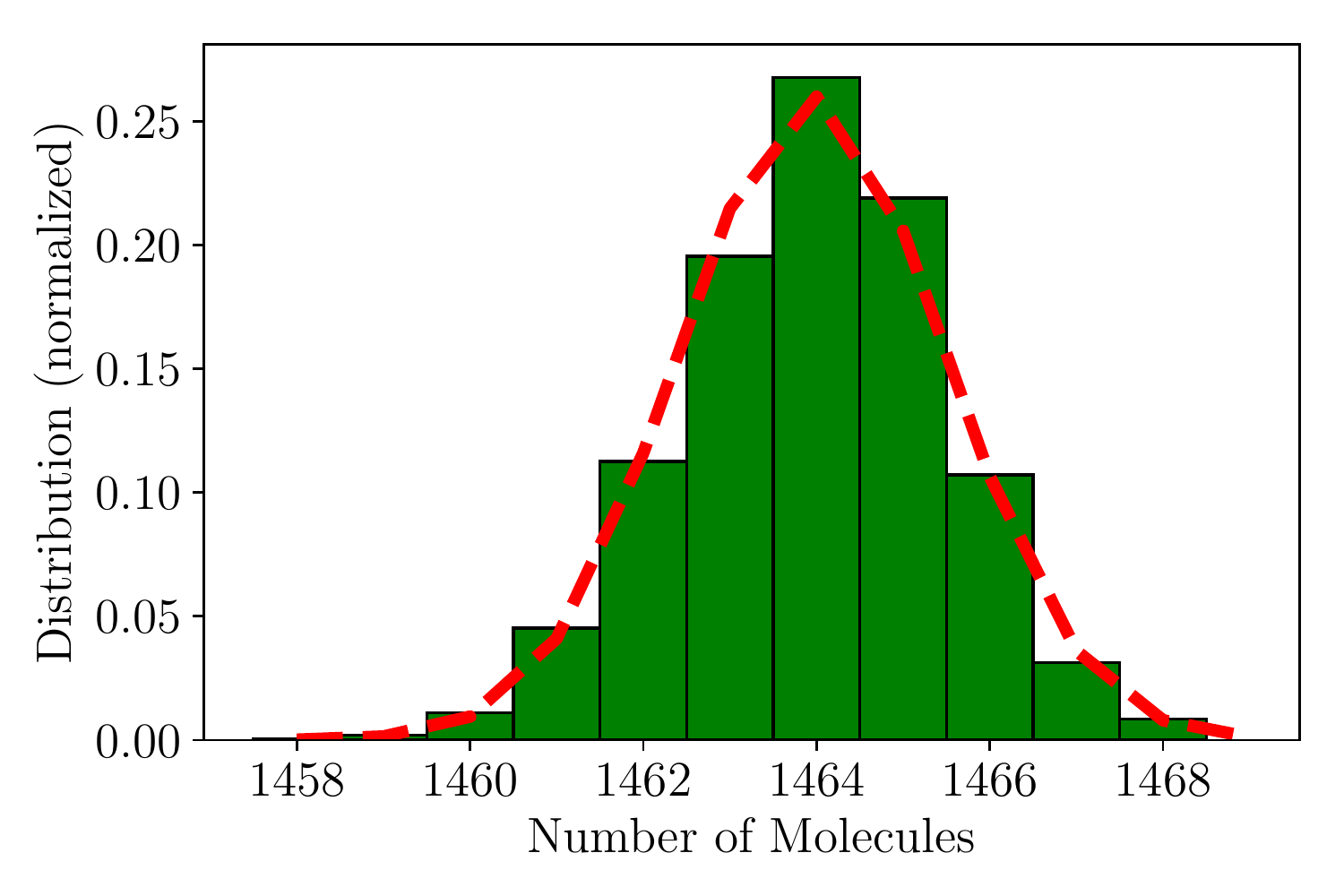}
\caption{ (Top) 5~nm spherical pore (left) and 40$\sigma$ length cylindrical pore (right). (Bottom) 2~nm spherical (left) and 80$\sigma$ length cylindrical pore (right). The fluctuation in the number of molecules at $T = 87.3$~K in both of the 5~nm pores and the 2~nm cylindrical pore exhibit a Gaussian distribution, whereas for the 2~nm spherical pore, the fluctuations do not fit well into a Gaussian distribution. The cylindrical pores can fit many more molecules, allowing larger fluctuations.}
\label{fig:hist} 
\end{figure}

The isothermal modulus is calculated based on the fluctuation of the number of atoms in the pores, so it is worth looking at the histograms for the systems giving such drastically different elastic moduli. Figure \ref{fig:hist} gives the histograms for argon atoms in four systems: 5~nm spherical and cylindrical pores and 2~nm spherical and cylindrical pores. While the 5~nm pore systems of both morphologies exhibit normally-distributed fluctuations in $N$, the 2~nm spherical micropores do not. The other mesopores not shown here were also normally distributed. The problem with the distribution in the 2~nm spherical pore is related to the smallness of the fluid system: the mean number of atoms in this pore is only around 60. The 2~nm cylindrical pore can be made arbitrarily long so that the number of atoms is sufficient to have normal distribution. Nevertheless, the cylindrical 2~nm pore does not provide reasonable values for the modulus. This is discussed below in Section \ref{sec:Discussion}.

\begin{figure}[H]
\centering
\includegraphics[width=0.7\linewidth]{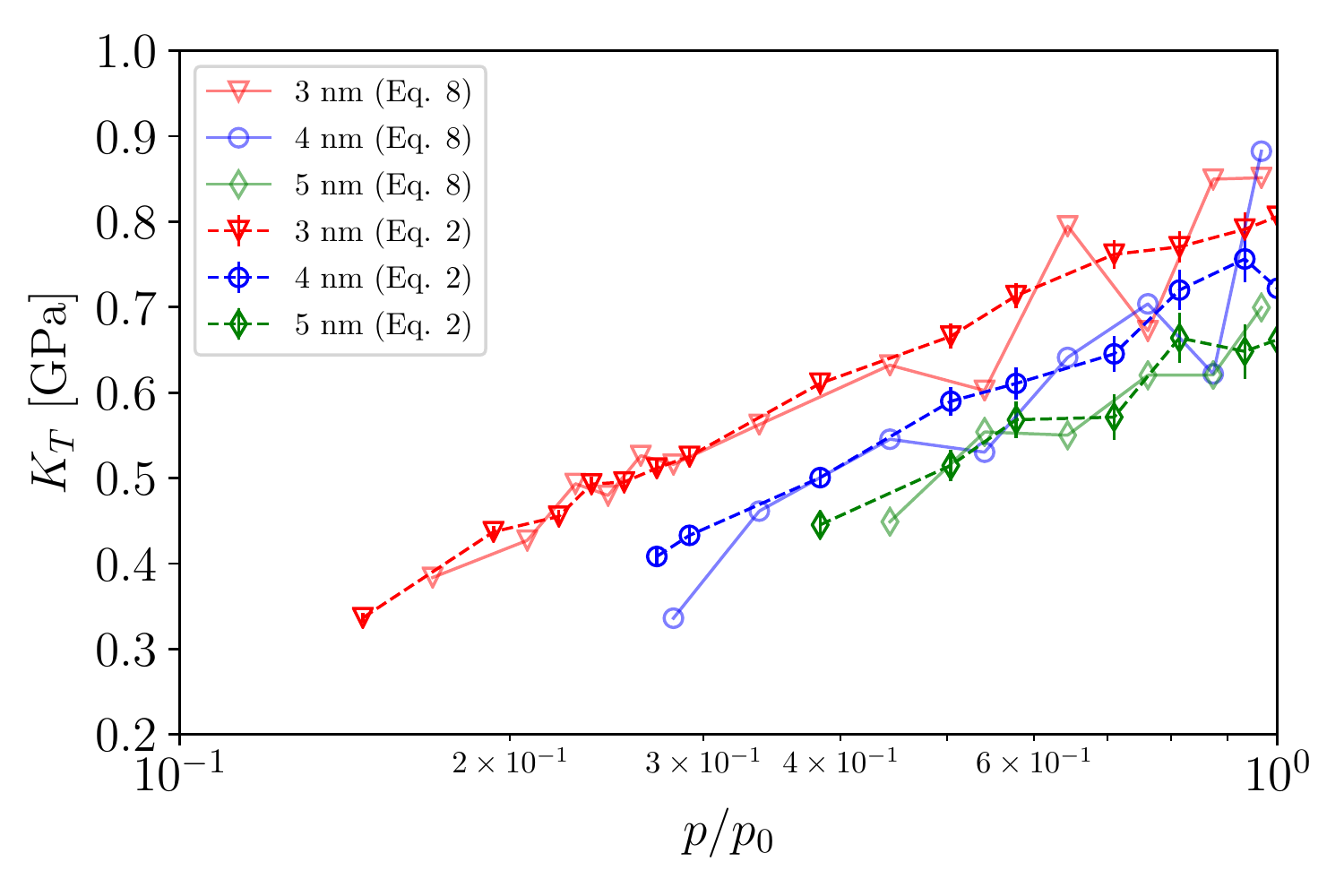}
\caption{The isothermal modulus at $T = 87.3$~K calculated using Eq. \ref{beta-thermo-final}, i.e. from the slope of the adsorption isotherm, along with the modulus calculated using Eq.~\ref{beta-fluct} for confined argon in spherical pores of 3, 4, and 5~nm pore width.}
\label{fig:K-thermo}
\end{figure}

The slopes of the isotherms in Figure \ref{fig:Isotherms} along the filled pore region allows for the calculation of the compressibilities (or elastic moduli) by Eq. \ref{beta-thermo-final}.  Figure \ref{fig:K-thermo} shows the isothermal modulus of the fluid in spherical pores of three sizes as a function of vapor pressure $p/p_0$ calculated based on two different methods: the method based on statistical mechanics (Eq. \ref{beta-fluct}) and the macroscopic method (Eq.~\ref{beta-thermo-final}). Notably, although the methods are very different, they produce very similar results. We do not show the calculations based on Eq.~\ref{beta-thermo-final} for cylindrical pores because of the scattered points on the isotherms for those systems. The application of Eq.~\ref{beta-thermo-final} for the calculation of the fluid modulus in cylindrical pores is discussed below for simulations at higher temperature.

\begin{figure}[H]
\centering
\includegraphics[width=0.7\linewidth]{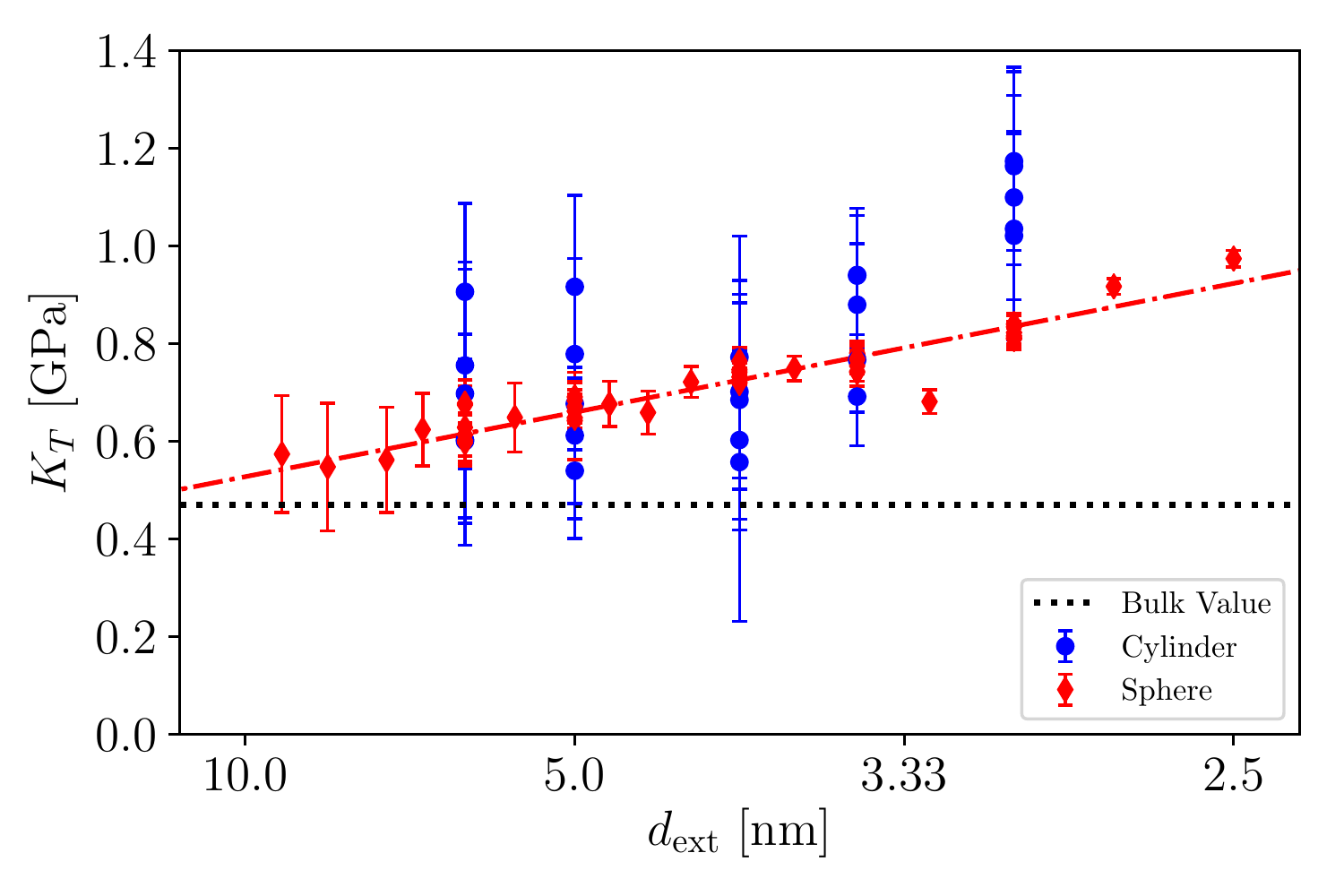} \\
\includegraphics[width=0.7\linewidth]{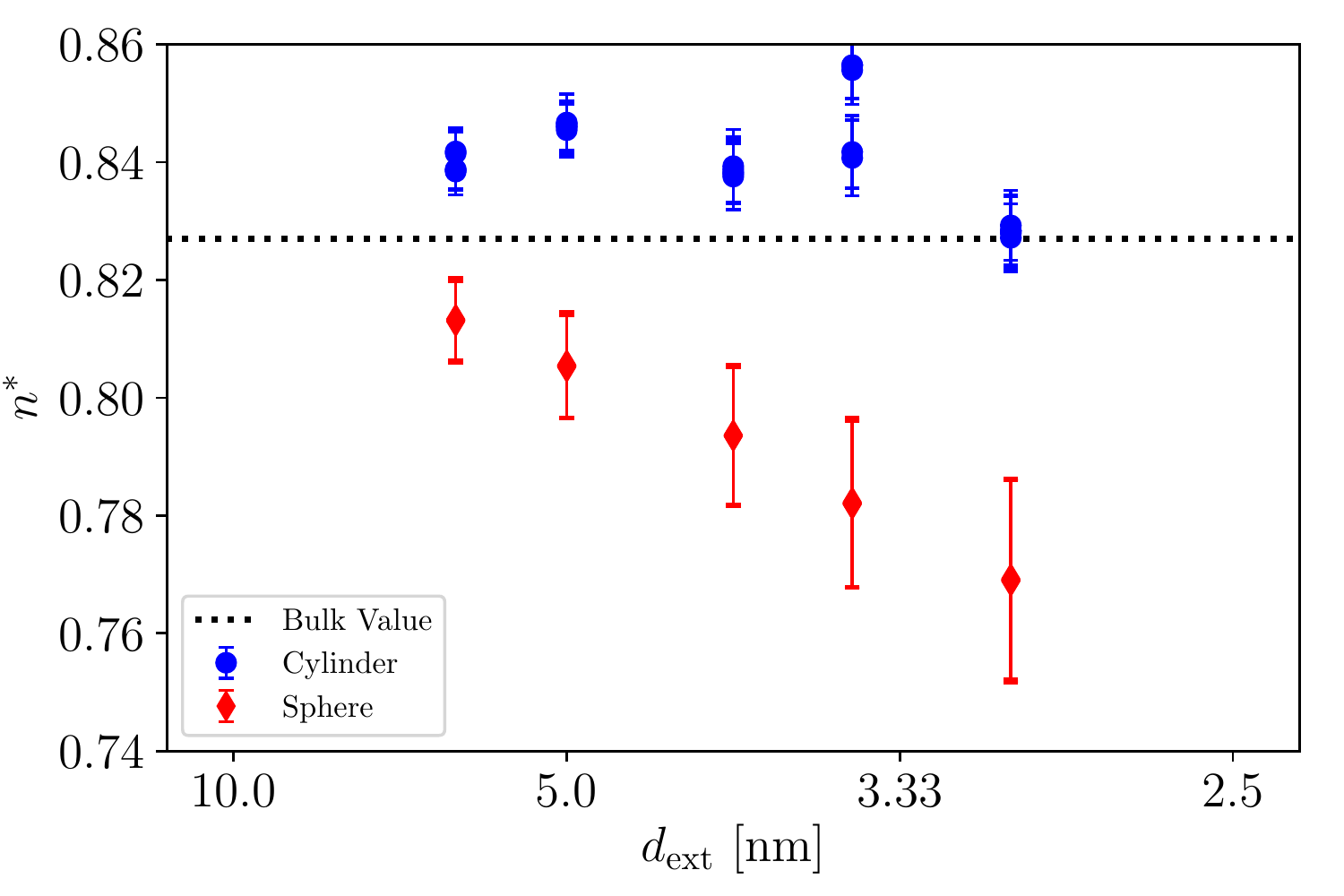}
\caption{(Top) Isothermal elastic modulus $K_T$ of argon in spherical and cylindrical pores at saturation pressure $p/p_0=1$ and $87.3$~K as a function of reciprocal pore size $1/d_{\rm{ext}}$ for $d_{\rm{ext}}$ values of 3, 3.5, 4, 5, and 6~nm. Also included are the data for spherical pores from Ref.~\cite{Gor2017Biot} covering the range of pore sizes between 2.5 and 9.0~nm. Error bars are correlation error estimated by the method described in Ref.~\onlinecite{Gor2015compr}. The dotted line shows the isothermal elastic modulus for the bulk liquid argon at saturation at $T=87.3$~K \cite{Tegeler1999}. (Bottom) Average fluid density at saturation point in a spherical and cylindrical pore as a function of the pore size. The dotted line shows the density of bulk fluid at the same thermodynamic conditions ($\mu$ and $T$). We find that the density of the cylindrical pores is higher than for the spherical pores, even though the well of the spherical interaction potential is deeper than for cylinders, as shown in Figure \ref{fig:Usf}}
\label{fig:K-density}
\end{figure}

The upper panel of Figure~\ref{fig:K-density} displays the isothermal modulus of a fluid at full saturation $p = p_0$ plotted as a function of the pore size. For spherical pores, the results obtained here agree well with data from Refs.~\onlinecite{Gor2015compr, Gor2017Biot}. These points show a linear dependence, approaching the bulk modulus value of $0.47$~GPa \cite{Tegeler1999} as the pores get larger than $10$~nm. The cylindrical pores show the modulus which is close to the spherical one, yet the significant scatter in the data does not allow to draw a conclusion about the trend.

The lower panel of Figure \ref{fig:K-density} displays the average fluid density in the pore at saturation as a function of reciprocal pore size. The density of bulk liquid argon at 87.3~K is 1.395 g/cm$^3$, which corresponds to a LJ reduced density of about $n_{\mathrm{bulk}}^*=0.827$ \cite{Tegeler1999}, is shown with the dotted line. There are two pronounced trends seen in this figure. First, the density of the confined fluid in spherical pores is lower than the bulk density, and it increases with the pore size trending towards the bulk value as the pore size increases above around 10~nm. The second trend is that for the same pore size, the density of fluid in cylindrical pores exceeds that in the spherical pores. Both of these trends were earlier discussed by Keffer et al. \cite{Keffer1996} for a LJ fluid in smaller pores. Recently densification of LJ fluid in the cylindrical pores was studied by Siderius et al. \cite{Siderius2017}, they found that the higher density of fluid in the cylindrical pores is related to its ordering, which does not take place in the bulk.

\begin{figure}[H]
\begin{center}
\includegraphics[width=0.7\linewidth]{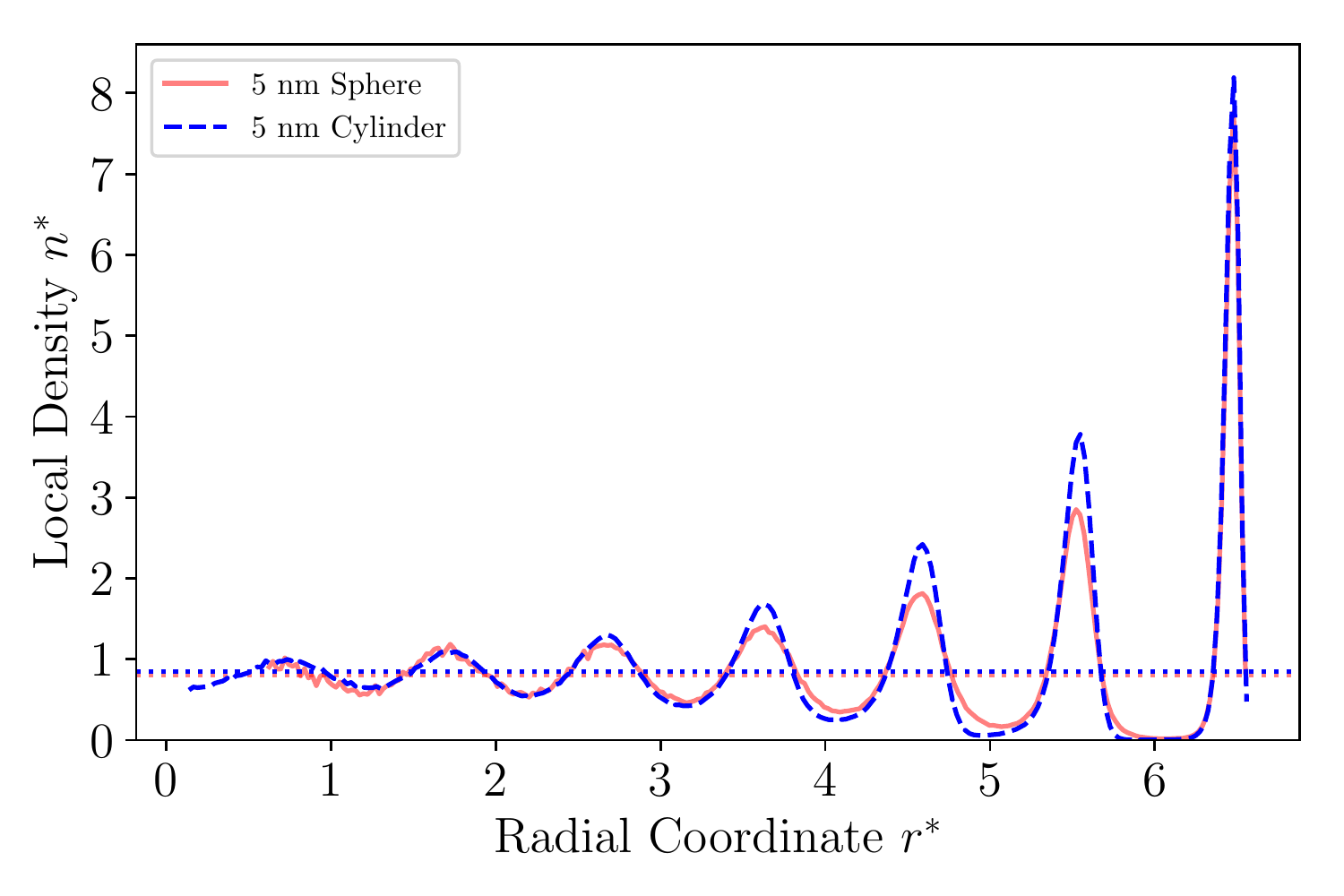}
\end{center}
\caption{Density profiles for argon confined in 5~nm spherical (solid red line) and cylindrical (dashed blue line) pores at saturation ($T = 87.3$~K, $\mu^{*} = -9.6$). The dotted lines of corresponding colors represent the average densities for each of the systems. The first several density peaks and wells near the adsorbing wall are more pronounced for the cylindrical pore than for the spherical one. This suggests that the cylindrical pore has a more ordered fluid phase.}
\label{fig:profiles} 
\end{figure}

The density of the fluid confined in a nanopore is not uniform. The interaction of the fluid atoms with the solid wall significantly alters the density and leads to the appearance of the dense layers in the vicinity of the solid walls. The density profiles for 5~nm pores of both spherical and cylindrical geometry are shown in Figure~\ref{fig:profiles}. These density profiles can explain some of the effects we observed for the elastic modulus $K_T$; a more detailed discussion is given in Section \ref{sec:Discussion}.

\begin{figure}[H]
\centering
\includegraphics[width=0.7\linewidth]{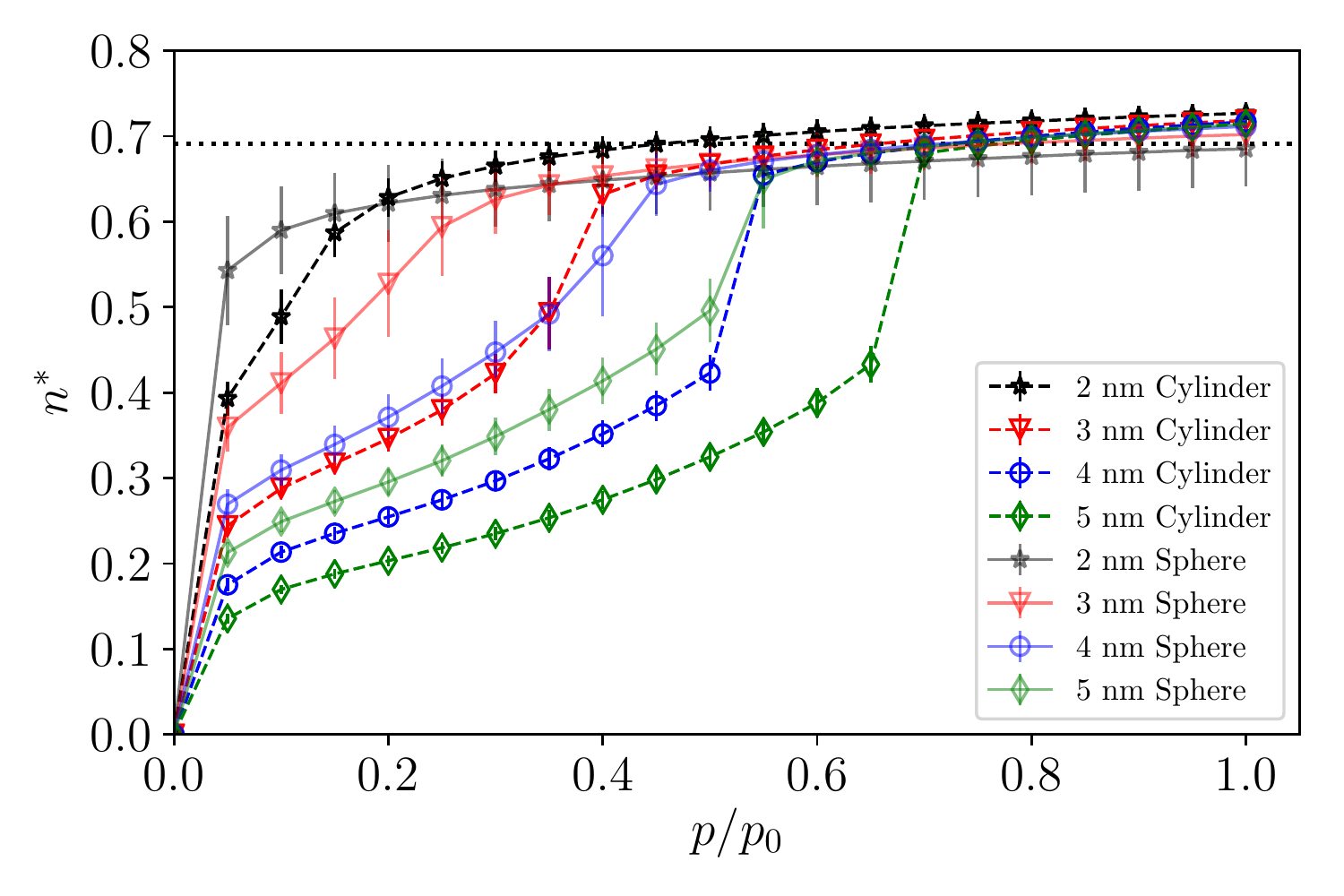}
\caption{GCMC adsorption isotherms for argon at 119.6~K in spherical and cylindrical pores shown as the average reduced fluid density $n^*$ plotted versus relative pressure. The horizontal dotted line at $n^* = 0.691$ represents the bulk density \cite{Tegeler1999}.}
\label{fig:Isotherms-120K} 
\end{figure}

Since our simulations of argon in cylindrical pores lead to inconclusive results on the elastic modulus, we ran additional simulations at higher temperature. We chose $T = 119.6$~K (corresponding to $T^* = 1$), which is noticeably higher than the normal boiling point of argon, yet is still far from the critical point ($T = 150.7$~K). Figure~\ref{fig:Isotherms-120K} shows the adsorption isotherms for cylindrical ($40 \sigma_{\rm{ff}}$ length) and spherical pores with sizes 2, 3, 4, and 5~nm.

\begin{figure}[H]
\centering
\includegraphics[width=0.7\linewidth]{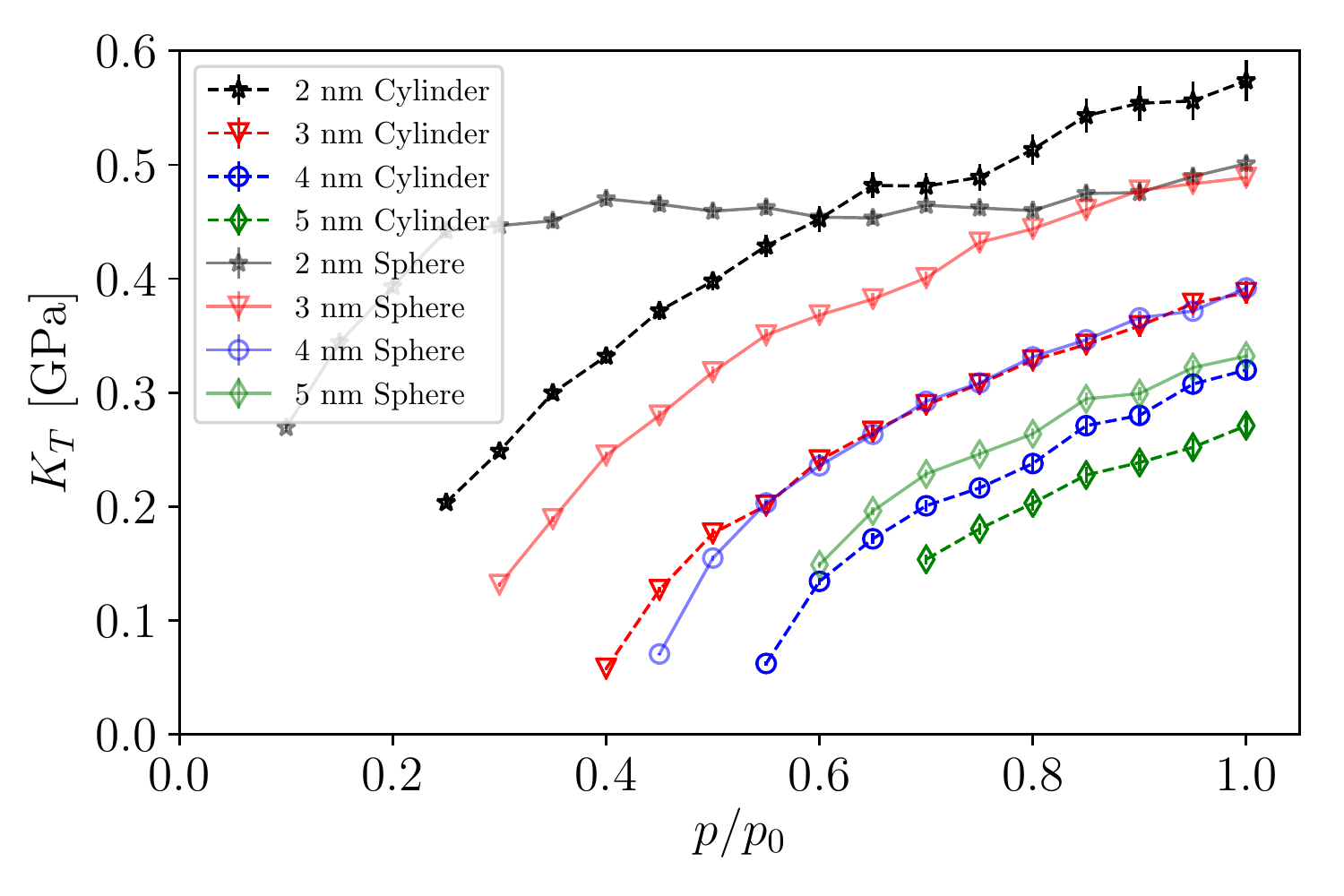}
\caption{The isothermal modulus $K_T$ of argon in spherical and cylindrical pores of 2, 3, 4, and 5~nm at $119.6$~K calculated using Eq. \ref{beta-fluct} from GCMC simulations as a function of reduced pressure. }
\label{fig:K-120K} 
\end{figure}

Similarly to the data at $T = 87.3$~K we carried out the calculation of isothermal elastic modulus based on the fluctuations of number of atoms in the pores (Eq.~\ref{beta-fluct}). The resulting curves for the moduli as a function of vapor pressure $p/p_0$ are shown in Figure~\ref{fig:K-120K}. The curves for both spherical and cylindrical pores show the clear monotonic trend, observed above in Figure~\ref{fig:K-all-points} for the data in the spherical pores. The only curve showing a less pronounced increase of the elastic modulus with the vapor pressure is for the 2~nm spherical pore, which, as discussed above, might be too small for application of our method.

\begin{figure}[H]
\centering
\includegraphics[width=0.7\linewidth]{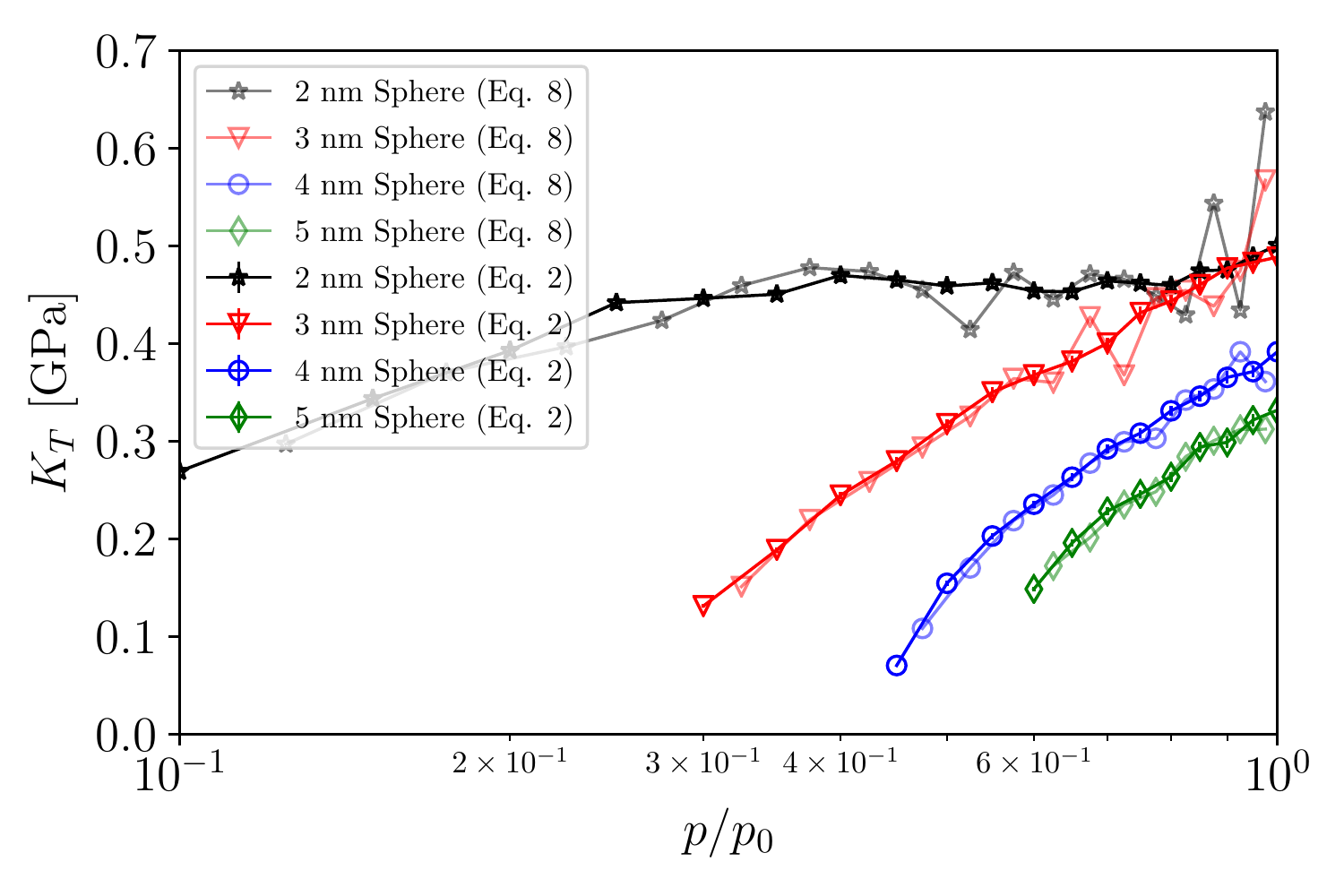}
\includegraphics[width=0.7\linewidth]{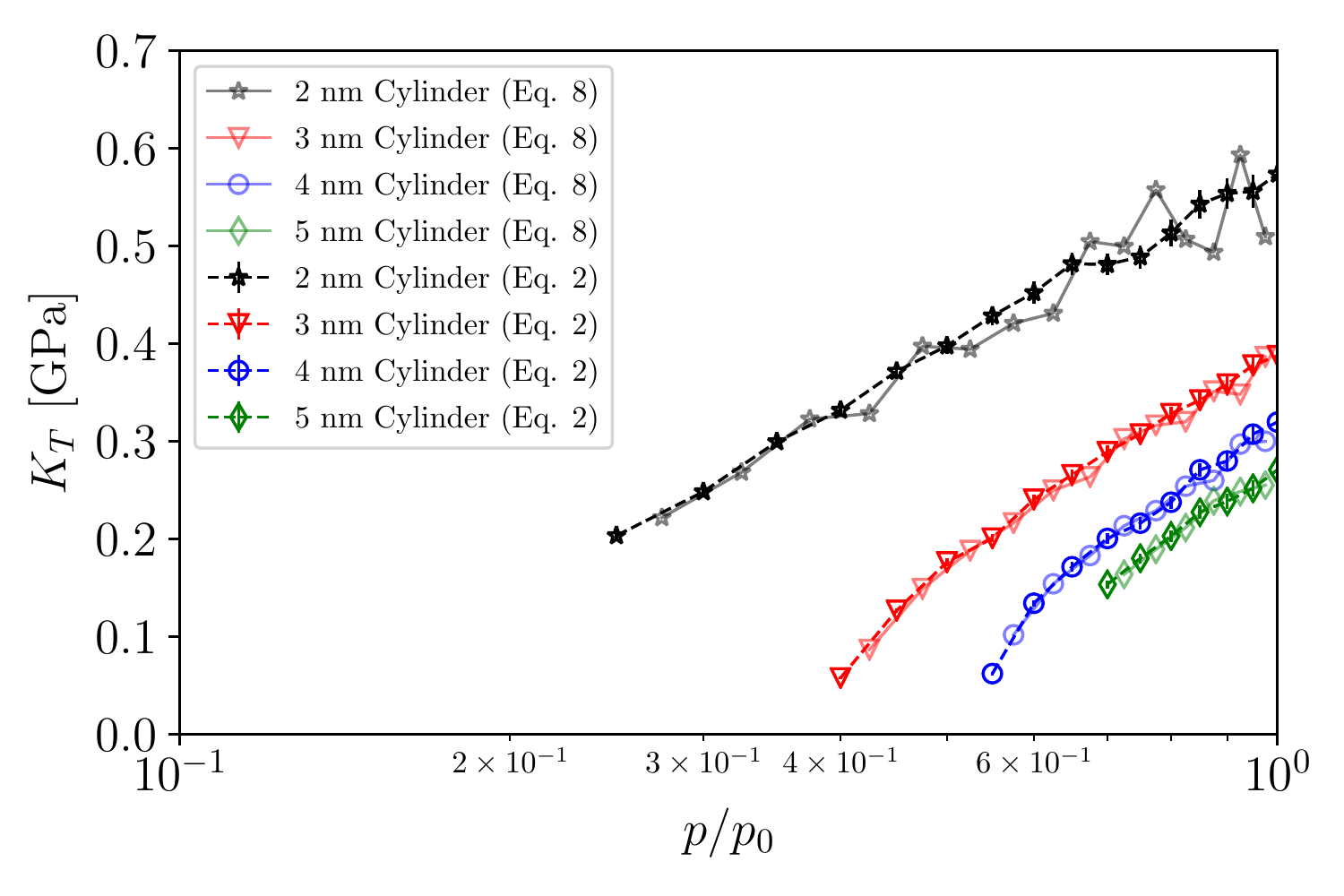}
\caption{The isothermal modulus of confined argon at $T = 119.6$~K calculated using Eq. \ref{beta-thermo-final}, i.e. from the slope of the adsorption isotherm, along with the modulus calculated using Eq.~\ref{beta-fluct} for confined argon in spherical (top) and cylindrical (bottom) pores of 2, 3, 4, and 5~nm diameter.}
\label{fig:K-thermo-120}
\end{figure}

Figure~\ref{fig:K-thermo-120} shows the isothermal moduli calculated for argon at $T =119.6$~K in spherical (top panel) and cylindrical (bottom panel) pores using the two different methods: the fluctuation method (Eq.~\ref{beta-fluct}) and thermodynamic method (Eq.~\ref{beta-thermo-final}). Similarly to Figure~\ref{fig:K-thermo} showing the agreement between the two methods for calculating the modulus for the simulation data for spherical pores at $T = 87.3$~K, Figure~\ref{fig:K-thermo-120} suggests that the two methods are fully consistent. Note that all of the data series, except for the curve for the 2~nm spherical pore, show a logarithmic dependence of the modulus on the vapor pressure. 

\begin{figure}[H]
\centering
\includegraphics[width=0.7\linewidth]{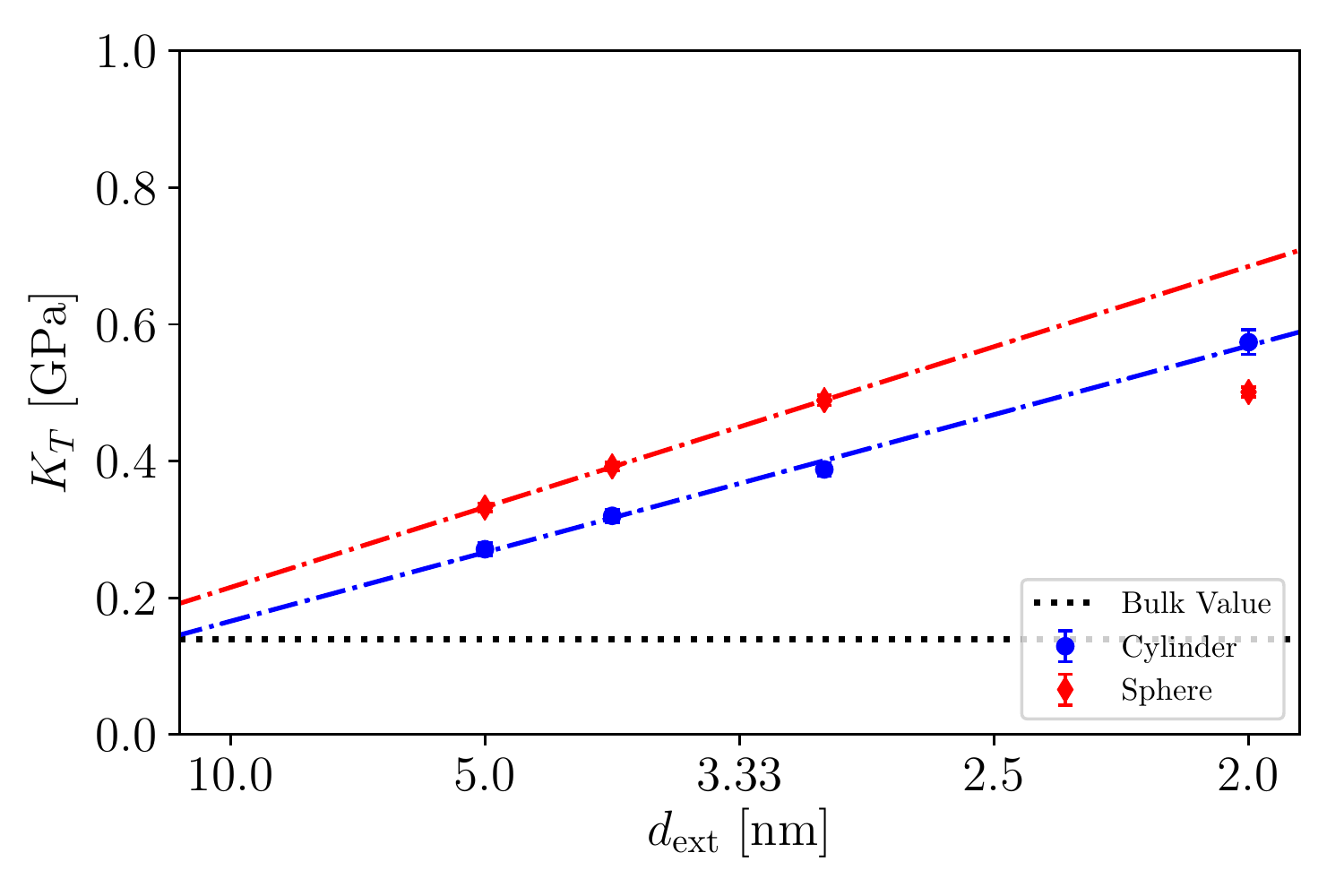} 
\caption{Isothermal elastic modulus $K_T$ of argon in spherical and cylindrical pores at saturation pressure $p/p_0=1$ and $T = 119.6$~K as a function of reciprocal pore size $1/d_{\rm{ext}}$ for $d_{\rm{ext}}$ values of 2, 3, 4, and 5~nm. The dash-dotted lines of the corresponding colors show the linear fit. The horizontal dotted line represents the bulk value of the elastic modulus.}
\label{fig:K-density-120}
\end{figure}

Figure \ref{fig:K-density-120}, similarly to Figure~\ref{fig:K-density}, shows the moduli at temperature $T = 119.6$~K at the saturation point ($p = p_0$). Unlike at the lower temperature, the elastic modulus of argon in cylindrical pores as a function of the pore size shows here the same linear monotonic trend as the modulus of argon in spherical pores. The linear fit is shown by the dash-dotted lines; the point corresponding to the 2~nm spherical pore is excluded from the linear fit.

\section{Discussion}
\label{sec:Discussion}

We presented the GCMC simulations of argon adsorption in spherical and cylindrical pores of various sizes. In addition to the adsorption isotherms, we calculated the elastic properties of adsorbed argon and compared the results for the two different pore morphologies. The main quantity we chose for consideration is isothermal (bulk) modulus of the fluid, a scalar thermodynamic property describing confined fluid as a macroscopic thermodynamic property. The rationale for introducing such modulus is driven by its accessibility in ultrasonic experiments \cite{Gor2018Gassmann}.

Since one of our central goals was to investigate the difference in elastic properties of the confined fluid, related to the morphologies of the confining pores, we started from the comparison of the solid-fluid interaction potentials $U_{\rm{sf}}$ for the spherical and cylindrical pores. We found that despite the difference in analytical forms for the integrated solid-fluid potentials for spherical and cylindrical pores, they have the same roots and therefore the internal diameters of cylindrical pores can be calculated from the external diameters using the same Eq.~\ref{dint}, which was initially written by Rasmussen et al. for spherical pores only \cite{Rasmussen2010, Gor2012}. Comparison of the depth of the potential wells for these two cases shows that the attractive potential for the spherical pore is stronger than for the cylindrical pore of the same size. This discrepancy between the potential depths explains the difference in the adsorption isotherm: capillary condensation in a spherical pore takes place at a lower pressure than in a cylindrical pore, in line with what has been discussed by Keffer et al. \cite{Keffer1996}.

\begin{figure}[H]
\centering
\includegraphics[width=0.7\linewidth]{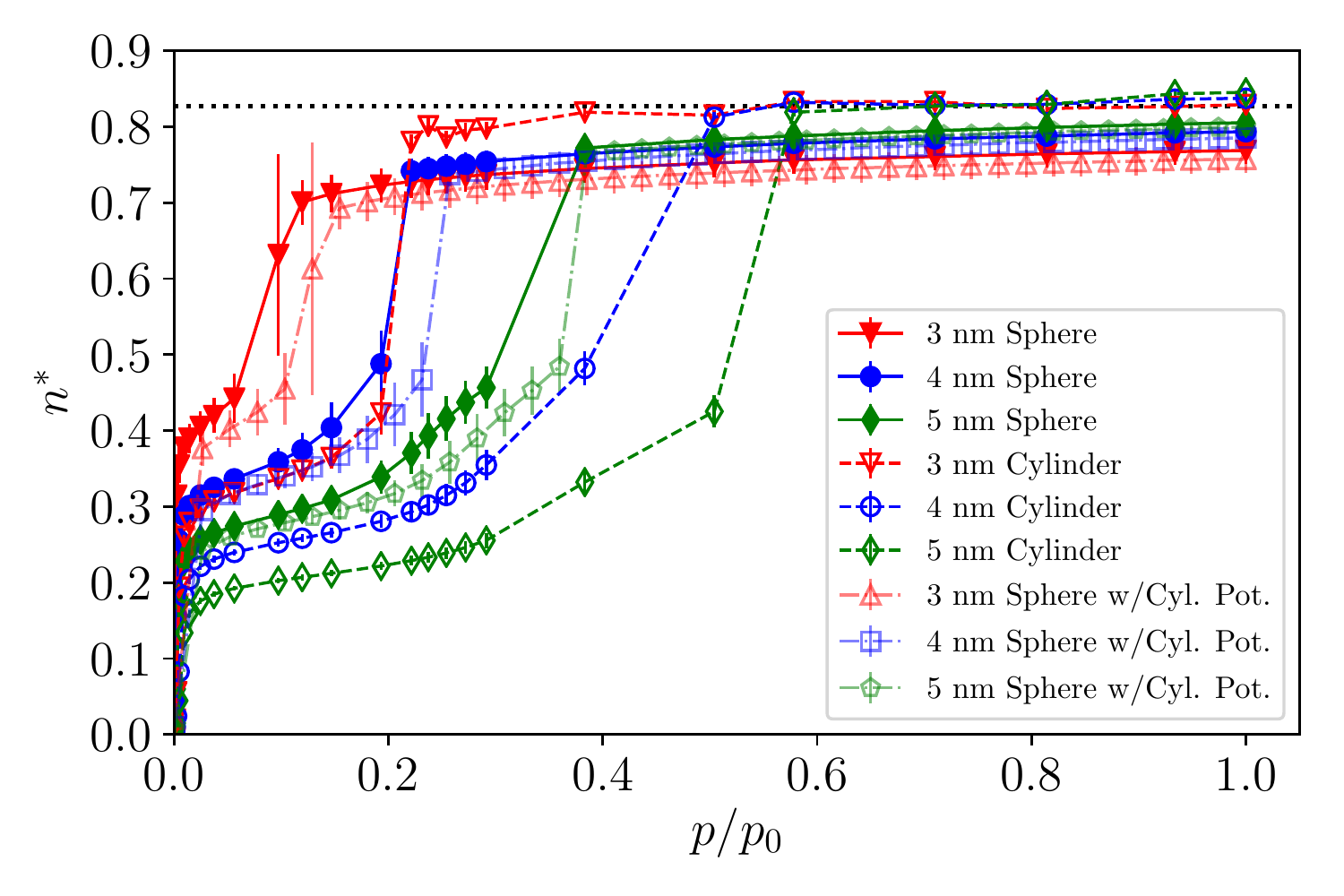}
\caption{GCMC adsorption isotherms for spherical, cylindrical and hybrid pores at $T= 87.3$~K shown as the average reduced fluid density $n^*=N\sigma_{\rm{ff}}^3/V$ plotted versus relative pressure. The horizontal dotted line displays the bulk density.}
\label{fig:Hybrid-Isotherms} 
\end{figure}

\begin{figure}[H]
\centering
\includegraphics[width=0.7\linewidth]{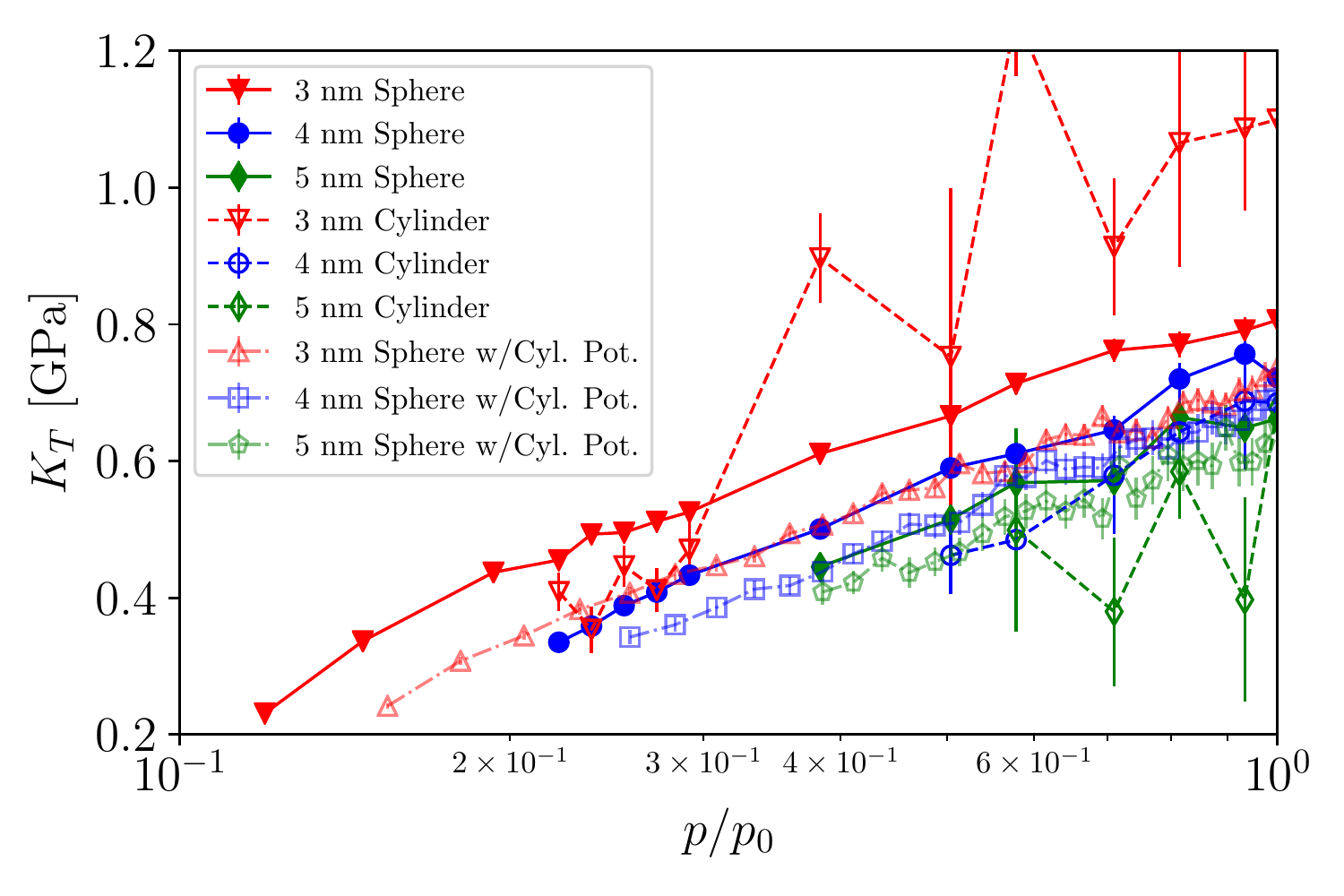}
\caption{Moduli calculated for argon confined in spherical, cylindrical and hybrid pores using Eq.~\ref{beta-fluct} at $T=87.3$~K, plotted versus relative pressure.}
\label{fig:Hybrid-Moduli} 
\end{figure}

Since spherical and cylindrical pore geometries are related to the two different solid-fluid interaction potentials, we carried out an additional test aiming to reveal the effect of the solid-fluid potential on the adsorption isotherms and elastic modulus. For this purpose we simulated the ``hybrid'' model: while using the spherical geometry we used the cylindrical potential. The resulting adsorption isotherms are shown in Figure \ref{fig:Hybrid-Isotherms}. The adsorption isotherms show that the capillary condensation in the spherical pores of each size take place at lower pressures than in hybrid model of the same size (which has the shallower potential well). The capillary condensation in the cylindrical pore takes place at even higher pressure than in the hybrid model, showing that the geometry itself reduces the confinement effects in addition to the solid-fluid interaction potential. Qualitatively similar effect of the solid-fluid potential on the elastic properties of the fluid is seen in Figure \ref{fig:Hybrid-Moduli}. The scatter on the modulus curve for the 4~nm cylindrical pore is relatively small, so this series can be compared to the two other models at the same pore size. The modulus of the fluid in cylindrical pore is lower than in hybrid model, and the hybrid is lower than in spherical. This trend is fully consistent with the trend for the capillary condensation point.

The calculated elastic moduli for various systems is the central part of this work. Our calculations of the elastic modulus for the confined fluid confirmed the trends reported earlier in Refs.~\onlinecite{Gor2015compr, Gor2016Tait, Gor2017Biot}. The first trend is the change in elastic modulus with the vapor pressure for each of the pore sizes, shown in the top panel of Figure \ref{fig:K-all-points} and \ref{fig:K-120K}. For both pore morphologies, there is a clear increase of the modulus with the increase of the relative vapor pressure $p/p_0$. The data for cylindrical pores at the normal boiling temperature are too scattered to make quantitative predictions, but for the spherical pores, it can be seen that the modulus changes as a logarithm of the vapor pressure. This logarithmic dependence has been observed in experiments \cite{Page1995, Schappert2014} and in molecular simulations using different techniques, DFT \cite{Gor2014} and transition-matrix Monte Carlo \cite{Gor2016Tait}. The origin of this dependence is the stretching of fluid by the negative Laplace pressure in the pore at $p < p_0$ \cite{Schappert2014, Gor2016Tait}. Note that in the simulations we do not model the curved liquid-vapor interface explicitly. Nevertheless, at any vapor pressure below $p_0$, the negative Laplace pressure acts on the fluid due to the difference in the chemical potential. The isothermal elastic moduli of many fluids display a linear dependence on pressure for a wide range of pressures (Tait-Murnaghan equation) \cite{Gor2016Tait, Hayward1967}: 
\begin{equation}
\label{Tait}
K_T(P)\simeq K_T(P_0) + K_T^{\prime} \cdot \left( P - P_0 \right)
\end{equation}
where, in our case, $P$ is the solvation pressure in the fluid phase, $P_0$ is some reference pressure, and $K_T^{\prime} = dK_T/dP$, which is constant in the first approximation. This dependence holds for confined fluids as well \cite{Gor2014, Keshavarzi2016}, moreover with the same slope $K_T^{\prime}$ \cite{Gor2016Tait}. The solvation pressure $P$ in the confined fluid (not to be confused with vapor pressure $p$) consists of two terms \cite{Gor2017review}: the solid-fluid interaction term and Laplace pressure
\begin{equation}
\label{Psolvation}
P = P_{\rm{sl}} + \frac{R_g T}{V_l} \ln \left( \frac{p}{p_0}\right).
\end{equation}
The logarithmic behavior of the second term in Eq.~\ref{Psolvation} together with Eq.~\ref{Tait} explains the logarithmic dependence of the fluid modulus on vapor pressure seen in Figures~\ref{fig:K-all-points} and \ref{fig:K-120K} and observed experimentally \cite{Page1995, Schappert2014}.

The second trend, which is clearly seen from the simulation data, is the dependence of the elastic modulus on the pore size, shown in the top panel of Figures~\ref{fig:K-density} and \ref{fig:K-density-120} for the modulus of fluid at saturation. For the spherical pores, as it was revealed earlier \cite{Gor2014}, the modulus $K_T$ is a linear function of the reciprocal pore size $1/d_{\rm{ext}}$. Similarly to the dependence on the vapor pressure, this dependence can be explained in terms of Tait-Murnaghan equation \ref{Tait} and the equation for solvation pressure \ref{Psolvation}. When the vapor is saturated ($p = p_0$) and the second term in Eq. \ref{Psolvation} vanishes, the pressure in the fluid is determined by the $P_{\rm{sl}}$ term, which scales like $1/d_{\rm{ext}}$ (Eq. 9 in Ref.~\onlinecite{Gor2017review}). Therefore, Eq. \ref{Tait} also gives the $1/d_{\rm{ext}}$ scaling for the elastic modulus $K_T$.

It is insightful to consider the trends observed for the moduli along with the trends for the fluid density. All the isotherms shown in Figure \ref{fig:Isotherms} display a well-known behavior: after capillary condensation, there is still a slow increase in the density of the fluid in the pore with the increase of vapor pressure $p/p_0$. This gradual densification of fluid in pores of all the sizes and morphologies corresponds to the gradual stiffening of the fluid -- an increase of its elastic modulus for each of the systems, as shown in Figure \ref{fig:K-all-points}. A similar comparison of the trends for densities and elastic moduli for different systems (pore sizes and morphologies) can be made based on the data shown in Figure \ref{fig:K-density}. The clear trend for modulus of the fluid in spherical pores corresponds to the clear trend for the density. This trend, however, is counter-intuitive: while the fluid is stiffer in smaller pores, its density in smaller pores is lower than in larger pores. This dependence of density of confined fluid on the pore size has been reported earlier in Ref.~\onlinecite{Keffer1996} and is related to the packing effects. 

The significant scatter in the results for argon modulus in cylindrical pores at 87.3~K complicated the comparison between the different pore morphologies. The likely reason for the scattering in the data for cylindrical pores is the layering of the fluid atoms along the straight pore walls. This layering causes the dense packing of the fluid in the pores. Figure \ref{fig:K-density} shows that the density of the fluid in cylindrical pores is noticeably higher than the density in spherical pores of the same size and even exceeds the bulk density. The dense packing in cylindrical pores makes the removal and insertion of atoms in GCMC very inefficient; therefore, the fluctuations of number of atoms in the pores may be lowered. Since the elastic modulus of the fluid is calculated based on the molecule fluctuations, lowering of the fluctuations will cause the apparent increase of the elastic modulus. This is indeed what we observed in our simulations for the smallest cylindrical pores: the fluid confined in a 2~nm cylindrical pore exhibited an extremely high elastic modulus (lower panel of Figure \ref{fig:K-all-points}). Such high values of the modulus exceed even the modulus of solid argon by an order of magnitude \cite{Stewart1968, Anderson1975, Utyuzh1983, Shimizu2001}, so it cannot be explained by the freezing of the fluid.

Simulations at higher temperatures, when Monte Carlo moves such as insertions and removals become more efficient, provide more reliable data for elastic moduli in cylindrical pores. As it is seen in Figures~\ref{fig:K-120K}, the simulation results at $T = 119.6$~K fall on smooth curves, which are suitable for comparison between spherical and cylindrical confinement. Moreover, the adsorption isotherms at $T = 119.6$~K in both pore geometries are sufficiently smooth for the calculation of the modulus using the thermodynamic route, i.e. by numerical differentiation of adsorption isotherms. Figure~\ref{fig:K-thermo-120} shows that at a higher temperature, the predictions of two methods for calculation of elastic modulus match perfectly.

Unfortunately, we should conclude that at the normal boiling point of argon, the pores of 2~nm in diameter and smaller (i.e. micropores) remain challenging irrespective of their morphology. The calculation of the fluid modulus in the 2~nm spherical pore was not feasible because of the smallness of the system: the average number of atoms at saturation pressure is ca. 60 and the fluctuations are not normally distributed (see the top right panel in Figure~\ref{fig:hist}). The cylindrical pore can be made arbitrarily long, so that the number of atoms at saturation in cylindrical pores is sufficiently large. For the 2~nm cylindrical pore of $L = 80 \sigma$ the average number of atoms is ca. 1460, and the fluctuations are normally distributed. Nevertheless, the strong attractive potential of the narrow confinement makes the GCMC insertions and removals very inefficient, so that the fluctuations are damped. These damped fluctuations result in an apparent high modulus, which is a computational artifact rather than the real behavior. However, at higher temperature, while the 2~nm spherical pore still remains a challenge, the calculation of the modulus in the 2~nm cylindrical pore gives results similar to other pore sizes. 

Both the high scattering in the results for the modulus of the fluid in cylindrical pores and the apparent high modulus in the smallest pores, suggest that while the model for cylindrical pores used here is suitable for calculation of adsorption isotherms, it is not efficient for the calculation of the elastic modulus at the temperature typically used in argon adsorption experiments. It is likely that the main drawback of the model is the smooth structureless cylindrical pore wall, which stimulates the fluid atoms to arrange in tightly packed layers along it (see Figure~\ref{fig:profiles}). There could be two possible solutions to this problem. The first solution is to consider the pores with atomistic details, representing molecularly rough surfaces of real amorphous materials, e.g. mesoporous silicas or Vycor glass. This approach will require explicit modeling of the solid atoms, increasing the computational cost. The second solution is to use one of the approaches that take into account the heterogeneity or molecular roughness of the pore walls, yet do not explicitly mimic the atomistic structure of the walls. Among such approaches, the two versions of DFT could be mentioned: the quenched solid DFT by Ravikovitch and Neimark \cite{Ravikovitch2006} and two-dimensional DFT by Jagiello and Olivier \cite{Jagiello2013ads, Jagiello2013carbon}. Note that recent DFT calculations, showed that another thermodynamic property of confined fluid, the heat of adsorption has been shown to be strongly influenced by the degree of surface roughness \cite{Cimino2017}. Therefore, we expect that introducing the surface roughness in the calculation of elastic properties of confined fluids might have a noticeable effect.

Last but not least, our simulations show that the thermodynamic route for calculation of elastic properties is fully consistent with the fluctuation route. This justifies the earlier results obtained by one of us using the thermodynamic route, where the average density of the fluid was calculated based on the density functional theory \cite{Gor2014}. Those calculations showed in particular that the logarithmic dependence of the fluid modulus on vapor pressure is valid, even at vapor pressures above $p_0$.

\section{Conclusion}
\label{sec:Conclusion}

Fluids confined in nanopores exhibit properties different from the properties of the same fluids in bulk. In this paper we focused on exploring the elastic properties of confined fluids: isothermal compressibility or elastic modulus. We calculated the modulus of liquid argon at its normal boiling point ($T = 87.3$~K) adsorbed in model silica pores of two different morphologies and various sizes. The main goal was to investigate the effect of the pore morphology on the elastic properties of confined fluid. We used conventional Monte Carlo simulations in the grand canonical ensemble to calculate argon adsorption isotherms for spherical and cylindrical pores with diameters of 2~nm and above. From the fluctuation of the number of fluid atoms in the pores at each given chemical potential, we calculated the elastic modulus of the fluid. Thus, for each of the considered systems, we obtained the elastic modulus as a function of vapor pressure. 

For spherical pores, for all the pore sizes exceeding 2~nm, we obtained a logarithmic dependence of fluid modulus on the vapor pressure. Calculation of modulus at saturation showed that the modulus of the fluid in the spherical pores is a linear function of the reciprocal pore size. The calculation of the modulus of the fluid in cylindrical pores appeared too scattered to make quantitative conclusions. Therefore, we performed additional simulations at higher temperature ($T = 119.6$~K), at which Monte Carlo insertions and removals become more efficient. The results of the simulations at higher temperature confirmed both regularities for cylindrical pores and showed quantitative difference between the fluid moduli in pores of different geometries. Both of the observed regularities for the modulus stem from the Tait-Murnaghan equation applied to a confined fluid. 

For the fluid in spherical pores at $T = 87.3$~K and for the fluid in both geometries at $T = 119.6$~K, we calculated the elastic moduli from the numerical differentiation of adsorption isotherms and the results appeared very close to the method based on the fluctuations of number of atoms. At the normal boiling temperature of argon, both methods of calculation of elastic modulus of the fluid showed themselves inefficient for pores of 2~nm (and smaller), therefore calculation of the elasticity of the fluid in micropores still remains a challenge, and will likely require the use of different simulation techniques.

Our results, along with the development of the effective medium theories for decoupling elastic properties in nanoporous systems, set the basis for analysis of the experimentally-measured elastic properties of fluid-saturated nanoporous materials. In particular, the relation between the pore size and the fluid modulus could serve as a groundwork for determination of pore sizes from the ultrasonic measurements. 

\section*{Appendix: Internal and External Diameters of Cylindrical Mesopores}

\begin{figure}[H]
\centering
\includegraphics[height=2.5in]{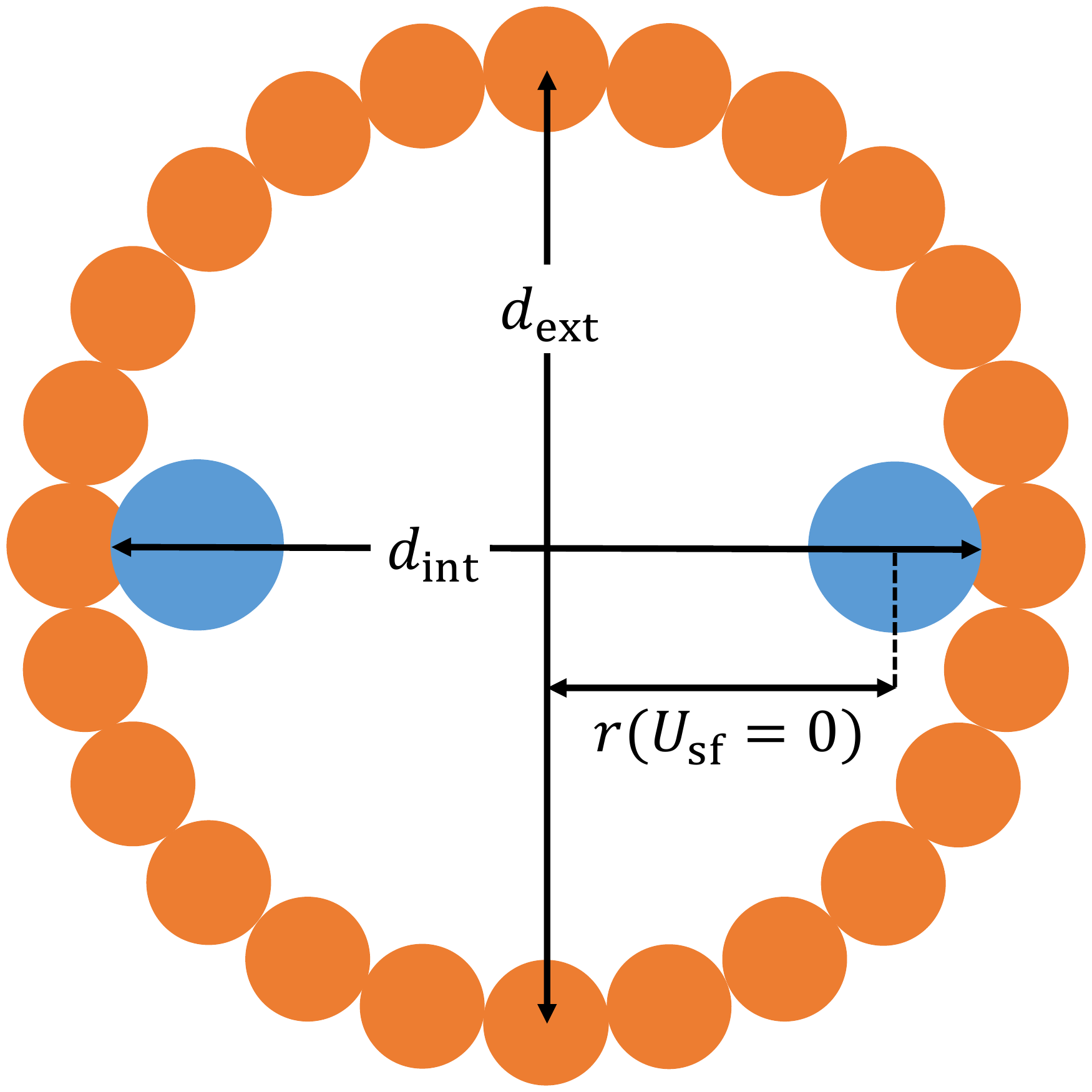}
\includegraphics[height=2.5in]{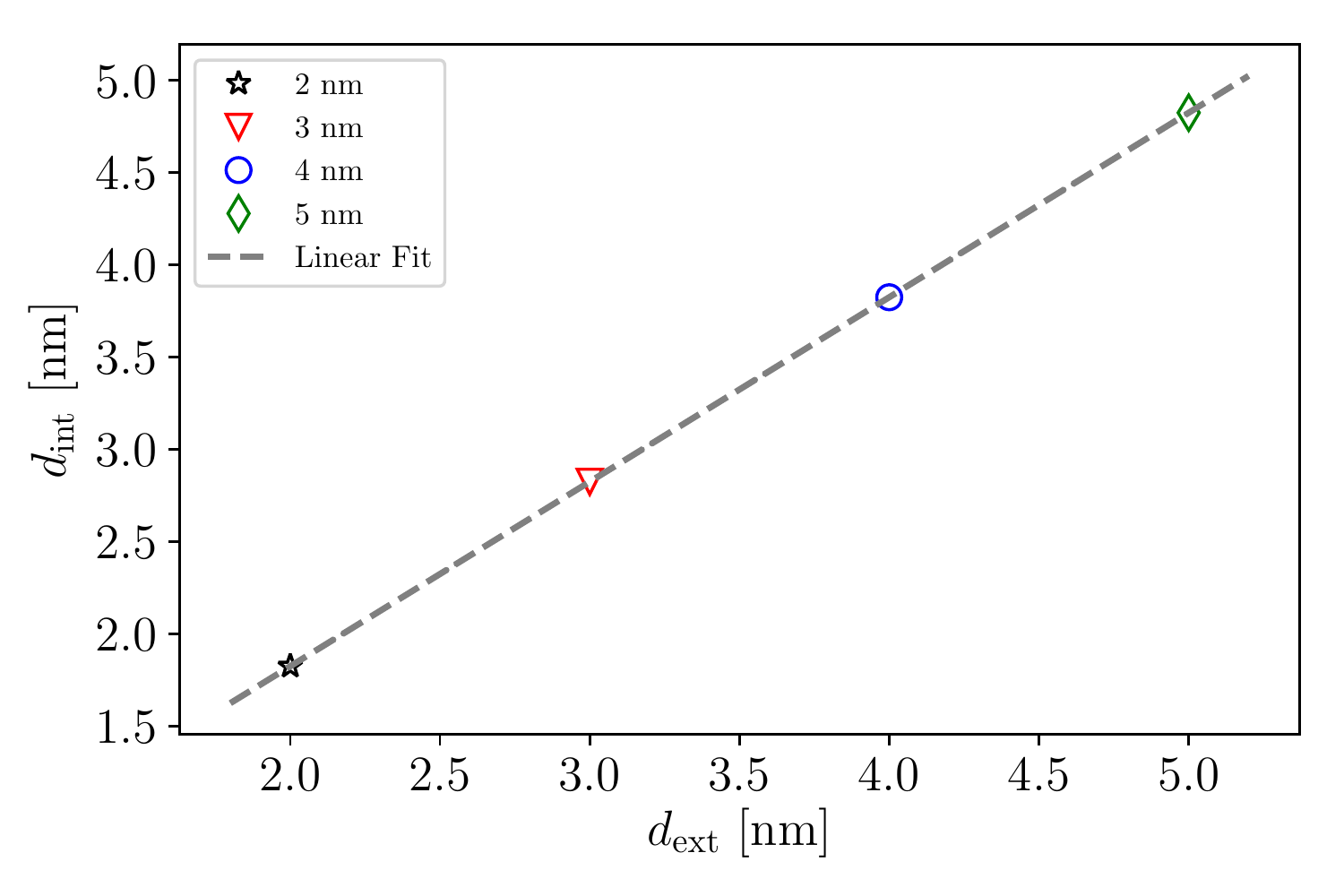}
\caption{(Left) Schematic of a cylindrical pore, showing the distinction between the internal and external diameters. Also noted is the radial distance from the center corresponding to the zero of the solid-fluid potential. (Right) The internal diameter of the pore as a function of external pore diameter, the markers originate from the numerical solution of equation $U_{\rm{sf}}(r) = 0$, the dashed line is the linear fit, Eq.~\ref{dint}.}
\label{fig:dint} 
\end{figure}

The pore size refers to the external diameter $d_{\rm{ext}}$ which is taken as the center-to-center distance from one pore wall molecule to the molecule on the opposite side of the pore. Obviously, the volume of the pore that is accessible to the fluid atoms (internal volume) $V$ is different from the volume calculated using the external diameter of the pore. This internal volume can be calculated based on the positions of the outter-most fluid atoms in the pore. The center of such an atom corresponds to the zero of the integrated solid-fluid potential $U_{\rm{sf}}$ (shown in Figure \ref{fig:Usf}) \cite{Rasmussen2010}. Since the volume needs to be taken up to the outer edge of those fluid atoms, an additional $\sigma_{\rm{ff}}$ needs to be added to the distance between the centers of such atoms (Figure \ref{fig:dint}, top). 

The right panel of Figure \ref{fig:dint} plots the internal diameters $d_{\rm{int}}$ calculated from the root of $U_{\rm{sf}}$ for the cylindrical pores with $d_{\rm{ext}} = 2, 3, 4$, and $5$~nm. The dashed line is the linear fit, which provides Eq.~\ref{dint}. Note that this equation does not differ from the equation for spherical pores \cite{Rasmussen2010, Gor2012}. 

\end{document}